%% file: main.tex
\documentclass[lettersize,journal]{IEEEtran}
\usepackage{fancyhdr,setspace}
\usepackage{comment}
\usepackage{caption}
\usepackage{graphicx}
\usepackage{wrapfig}
\usepackage{url}
\usepackage{xurl}
\usepackage[compact]{titlesec} 
\usepackage{cite}
\usepackage{blindtext, graphicx, tabularx, multicol, lipsum,eqnarray,amsmath,chemformula,multirow, subfigure}
\usepackage{enumitem}
\usepackage{amsthm}
\usepackage[utf8]{inputenc}
\usepackage{amssymb,amsmath}
\usepackage[boxed]{algorithm2e}
\usepackage[noend]{algpseudocode}
\usepackage{float}
\algnewcommand{\IfThenElse}[3]{
  \State \algorithmicif\ #1\ \algorithmicthen\ #2\ \algorithmicelse\ #3}
\makeatletter
\newcommand{\removelatexerror}{\let\@latex@error\@gobble}
\makeatother
\usepackage{letltxmacro}
\newtheorem{definition}{Definition}

\ifCLASSINFOpdf

\else

\fi

\newcommand\blfootnote[1]{%
  \begingroup
  \renewcommand\thefootnote{}\footnote{#1}%
  \addtocounter{footnote}{-1}%
  \endgroup
}

\begin{document}
%
\title{Characterizing Coherent Integrated Photonic Neural Networks under Imperfections}



\author{Sanmitra Banerjee,~\IEEEmembership{Student Member,~IEEE}, Mahdi Nikdast,~\IEEEmembership{Senior Member,~IEEE}, and Krishnendu Chakrabarty,~\IEEEmembership{Fellow,~IEEE} 

\thanks{S. Banerjee and K. Chakrabarty are with the Department of Electrical and Computer Engineering at Duke University, Durham, NC.} 
\thanks{M. Nikdast is with the Department of Electrical and Computer Engineering at Colorado State University, Fort Collins, CO.}
\vspace{-3em}
}

\maketitle
\thispagestyle{plain}
\pagestyle{plain}
\begin{abstract}
Integrated photonic neural networks (IPNNs) are emerging as promising successors to conventional electronic AI accelerators as they offer substantial improvements in computing speed and energy efficiency. In particular, coherent IPNNs use arrays of Mach--Zehnder interferometers (MZIs) for unitary transformations to perform energy-efficient matrix-vector multiplication. However, the underlying MZI devices in IPNNs are susceptible to uncertainties stemming from optical lithographic variations and thermal crosstalk and can experience imprecisions due to non-uniform MZI insertion loss and quantization errors due to low-precision encoding in the tuned phase angles. In this paper, we, for the first time, systematically characterize the impact of such uncertainties and imprecisions (together referred to as imperfections) in IPNNs using a bottom-up approach. We show that their impact on IPNN accuracy can vary widely based on the tuned parameters (e.g., phase angles) of the affected components, their physical location, and the nature and distribution of the imperfections. To improve reliability measures, we identify critical IPNN building blocks that, under imperfections, can lead to catastrophic degradation in the classification accuracy. We show that under multiple simultaneous imperfections, the IPNN inferencing accuracy can degrade by up to 46\%, even when the imperfection parameters are restricted within a small range. Our results also indicate that the inferencing accuracy is sensitive to imperfections affecting the MZIs in the linear layers next to the input layer of the IPNN.
\end{abstract}




%
\IEEEpeerreviewmaketitle

\vspace{0.5em}
\section{Introduction}
\blfootnote{This work was supported in part by the National Science Foundation (NSF) under grant numbers CCF-1813370, CCF-2006788, and CNS-2046226.}
\label{intro}
\input{intro}

\section{Background and Related Prior Work}
\label{background}
\input{background}
\section{Uncertainties in Phase Angles and Splitting Ratios: A Bottom-Up Characterization}
\label{random}

\input{random}

\section{Optical Loss and Quantization Errors}
\label{imprecision}
\input{systematic}

\subsection{IPNNs with Low-Precision Phase Encoding}

\input{precision}
\section{Case Study: IPNNs under Simultaneous Uncertainties and Imprecisions}
\label{case_study}
\input{case_study}

\section{Conclusion}

We have presented the first comprehensive characterization of the inferencing accuracy of a trained IPNN under imperfections that arise from several sources, including spatially correlated and localized uncertainties in phase angles and splitting ratios, non-uniform MZI insertion loss, and low-precision phase encoding. We have analyzed these imperfections in a bottom-up approach where uncertainties are introduced in the optical components and their impact is observed at the component, device, layer, and system level. Considering different uncertainty scenarios, we have demonstrated that the degradation in the IPNN accuracy due to an uncertainty depends not only on the magnitude and distribution of the uncertainties but also on the position of the affected component. Based on the extensive simulations in our study, the main observations that can guide the development of reliable IPNNs are as follows:  

\begin{itemize}
    \item The inferencing accuracy loss due to multiple simultaneous imperfections is lower than the sum of the accuracy losses due to each standalone uncertainty -- \textit{the individual effects of IPNN imperfections are not additive};
    \item Under similar levels of uncertainties in phase shifters and beam splitters (standard deviations $\sigma_{PhS}$ and $\sigma_{BeS}$ remaining constant), the IPNN accuracy decreases with increasing the correlation length ($L$) in uncertainties;
    \item Uncorrelated uncertainties in PhS and BeS with a large $\sigma_{PhS}$ and $\sigma_{BeS}$ are more catastrophic than  highly correlated variations with a small $\sigma_{PhS}$ and $\sigma_{BeS}$;
    \item Tuned phase angles should be encoded in memory using a precision of at least 7 bits;
    \item Deviation in the tuned phase angles (e.g., due to process variations, thermal crosstalk, and low-precision DACs) has a dominant impact among the IPNN imperfections;
    \item The IPNN accuracy is more sensitive to insertion loss and quantization error in the MZIs in the initial layers -- \textit{mitigative steps should focus on the initial IPNN layers}.
\end{itemize} 

We have also shown that the problem of finding the maximum acceptable limits of different imperfections to ensure a minimum IPNN accuracy is non-trivial and necessitates a computationally efficient search algorithm. \textcolor{black}{Based on simulation results, we have observed that the criticality of MZIs in an IPNN depends on their position and tuned phase angles. Consequently, for a different application (other than MNIST), the set of critical MZIs can vary. However, as all IPNNs are, fundamentally, cascaded linear multipliers, we expect that the trends of the impact of imperfections observed in this stuy will still hold. Also, the proposed analysis framework can be easily extended to additional datasets and design-time and run-time uncertainties and can help IPNN designers to model the performance of IPNNs under multiple simultaneous uncertainties and to develop efficient compensation methods during design time.}



\bibliographystyle{IEEEtran}
{
\hyphenpenalty=10000
\exhyphenpenalty=10000
\sloppy
\bibliography{Refs}
}

\appendix
\input{appendix.tex}

\end{document}

%% file: intro.tex
The rapid emergence of big data from mobile, Internet of Things (IoT), and edge devices and the continuous growth in computing power have enabled deep neural networks (DNNs) to perform various complex tasks, such as natural-language processing, action recognition, game-playing, and image classification \cite{raghu2020survey}. The primary computational primitive while querying such advanced DNNs is the time- and energy-intensive matrix multiplication operations \cite{denton2021acceleration}. To improve DNN efficiency, recent years have seen a push towards domain-specific artificial intelligence (AI) accelerators that use tightly coupled data processing units connected in a systolic array \cite{jouppi2018motivation}. However, with Moore's law approaching its end, electronic accelerators are facing fundamental bottlenecks due to the slowdown in CMOS scaling and low-bandwidth metallic interconnects \cite{waldrop2016chips}. Continued progress in AI development is also hampered by the high energy overhead associated with training and inferencing DNNs on electronic processors \cite{thompson2020computational}. \par

Integrated photonic neural networks (IPNNs) based on silicon photonics can expedite extensive linear operations (i.e., matrix multiplication) in DNNs \cite{sunny2021survey}. By taking advantage of the natural parallelism in photonics, computations in IPNNs can be performed in parallel, hence reducing the complexity of matrix-vector multiplication in DNNs from $O(N^2)$ to \textcolor{black}{approximately} $O(1)$ \cite{cheng2020silicon}. \textcolor{black}{Although the energy consumption in IPNNs increases with matrix sizes, simulation results have shown that optical matrix-vector multiplications can even outperform digital irreversible computation at the thermodynamic limit \cite{hamerly2019large}.} Moreover, experimental implementations of coherent IPNNs have demonstrated high accuracy, fast convergence during training, and the capability to learn non-linear decision boundaries \cite{zhang2021optical}. Such benefits along with continuous advances in CMOS-compatible silicon photonics technology have positioned IPNNs as a promising alternative to electronic AI accelerators \cite{cartlidge2020optical}.\par


Neurons in an IPNN can be implemented either in a coherent or a noncoherent manner. Coherent neurons use Mach--Zehnder interferometers (MZIs) with phase shifters (PhS) and 3-dB beam splitters (BeS) to modify the phase and amplitude of the single-wavelength optical signal \cite{sludds2020scalable}. The optical phase shift in such neurons is realized by inducing changes in the refractive index of the silicon (Si) waveguide using either the thermo-optic effect \cite{harris2014efficient}, electro-optic effect \cite{macik2017optimization}, or phase-change materials \cite{petri2018thermodynamic}. Fig. \ref{placeholder} presents a hierarchical view of a multi-layer perceptron (MLP)-based IPNN. On the other hand, noncoherent neurons employ the Broadcast-and-Weight configuration to manipulate the power of multiple-wavelength optical signals \cite{sunny2021survey}. While noncoherent neurons support wavelength-division multiplexing, they suffer from inter-channel crosstalk and the dependency between the input and output wavelengths necessitates energy-intensive wavelength conversion steps \cite{sunny2021survey}. As a result, recent commercial efforts on developing photonic neural networks have focused on using coherent neurons to expedite matrix multiplication at a low energy cost \cite{ramey2020silicon}. We explore IPNNs built using coherent neurons in this paper; after this point, the term IPNN, if not explicitly mentioned otherwise, denotes a coherent IPNN. 


Despite the aforementioned advances, there exist several roadblocks to the further advancement of IPNNs. In particular, their performance can be highly impacted by the optical losses accumulating when cascading MZI devices~\cite{clements2016optimal,reck1994experimental}, and the additional computation needed for mapping the trained weights---obtained during software training---to the parameters (i.e., phase angles) in MZI networks~\cite{clements2016optimal}. In addition, fabrication-process variations (FPVs) and mutual thermal crosstalk due to convective heat transfer between microheaters in thermo-optic PhS~\cite{jacques2019optimization} (considered in this paper) can lead to faulty matrix multiplication. FPVs arise from optical-lithography process non-idealities. For example, changes in the critical dimensions (e.g., waveguide width and thickness) lead to incorrect operation of photonic components~\cite{Mahdi_DATE16}. IPNNs are also prone to imprecisions caused by the non-uniform insertion loss of constituent MZIs \cite{shokraneh2020diamond}. In addition, the resolution of the phase settings depends on the encoding precision of the digital-to-analog converter (DAC). For example, using an $N$-bit DAC, only $2^N$ different phase angles in the range [0, 2$\pi$] can be realized. This leads to a quantization error in the encoded phase angles, which, in turn, results in a degraded inferencing accuracy. All these imperfections highly impact the performance and operation of IPNNs, and therefore must be fully characterized.\par


In this paper, we present the first comprehensive study of the impact of uncertainties and imprecisions (collectively referred to as imperfections) on the performance of IPNNs. Our analysis follows a bottom-up approach: we show how imperfections in different components affect the functionality of IPNN's fundamental devices (i.e., MZIs), and how the affected MZIs lead to incorrect matrix multiplication, which finally leads to a degraded inferencing accuracy. We show that this degradation varies significantly based on the tuned-phase angles, the position of the affected MZIs, and the nature of the variations (e.g., whether they are correlated or localized). Moreover, we show that imprecisions introduced due to non-uniform MZI insertion losses and quantization error in low-precision phase encoding have a catastrophic impact when topologically shallower (close to the input) IPNN layers are affected. In particular, we show that the IPNN accuracy can drop to 10\% (accuracy of random prediction for the MNIST dataset) under expected levels of uncertainties in PhS and BeS. The main contributions of this paper are:
\begin{figure}[t]
  \centering
  \includegraphics[width=0.48\textwidth]{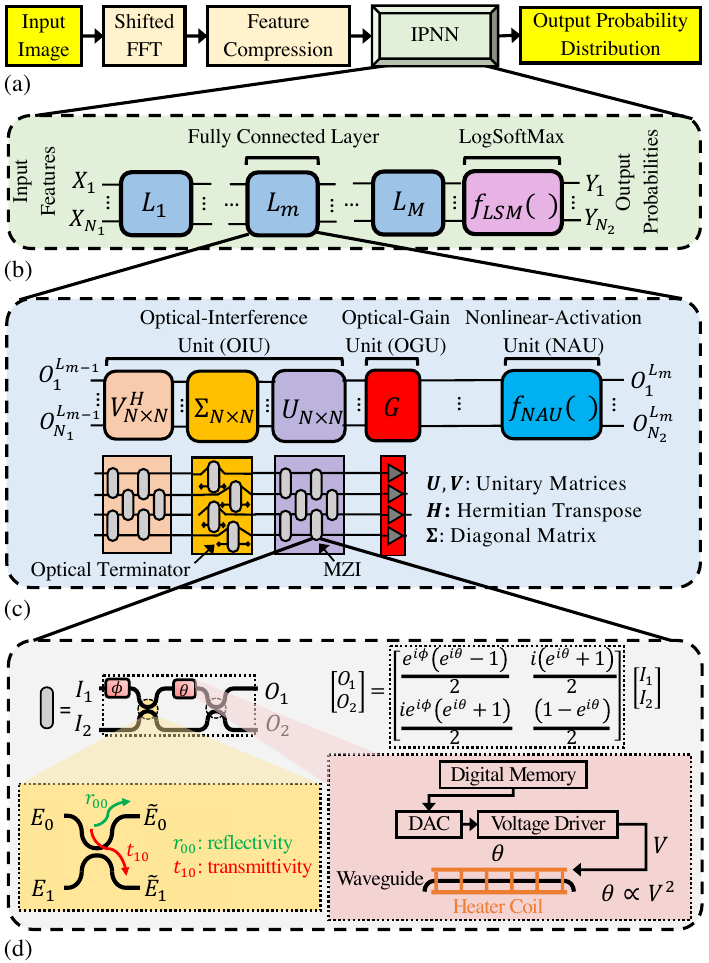}
  \caption{Hierarchical schematic of MLP-based coherent IPNNs.}
  \vspace{-1.8em}
  \label{placeholder}
\end{figure}
\begin{itemize}
    \item A comprehensive hierarchical analysis of the impact of uncertainties in phase angles and splitting ratios in the MZIs in IPNNs;
      \item An analysis of the impact of imprecisions due to the non-uniformity in the MZI insertion loss on IPNN accuracy;
      \item An analysis of IPNN performance under quantization errors in the PhS and a comparative study of different phase encoding methods;
      \item A modeling framework to identify critical components in IPNNs where imperfections lead to severe degradation in the IPNN inferencing accuracy;
    \item A case study on the inferencing accuracy of IPNNs in the presence of multiple simultaneous imperfections.
\end{itemize}


The remainder of this paper is organized as follows. Section II reviews the fundamentals of MZI-based coherent IPNNs, and discusses the different sources of IPNN imperfections (uncertainties and imprecisions) and prior work. In Section III, we analyze the impact of uncertainties in the phase angles and splitting ratios in IPNNs and identify the uncertainty susceptible ``critical" components in the network. Section IV focuses on the IPNN imprecisions; we first characterize the performance of IPNNs in the presence of lossy MZIs and then explore how the IPNN inference accuracy is degraded under low-precision phase settings. In Section V, we use the imperfection models proposed in Sections III--IV to demonstrate how the IPNN performance can be affected in the simultaneous presence of multiple imperfections. We conclude in Section VI by summarizing the observations that can guide the design and manufacturing of reliable next-generation IPNNs.

%% file: background.tex
\subsection{Mach--Zehnder Interferometer (MZI)}
As shown in Fig. \ref{placeholder}(d), an MZI consists of two tunable phase shifters (PhS, $\phi$ and $\theta$) on the upper arm and two 3-dB beam splitters (BeS). The PhS are used to apply configurable phase shifts and obtain varying degrees of interference between the input optical signals. They can be implemented using thermal microheaters, where the refractive index of the underlying waveguide changes with temperature (i.e., thermo-optic effect), altering the phase of the optical signal traversing the waveguide. Moreover, 2$\times$2 BeS can be designed using directional couplers, where a fraction (defined by transmittance) of the optical signal at an input port is transmitted to an output port, and the remaining (defined by the reflectance) is coupled to the other output port with a phase shift of $\frac{\pi}{2}$. For symmetric 3-dB (i.e., 50:50) BeS, both transmittance and reflectance coefficients are $\frac{1}{\sqrt{2}}$. As a result, the transfer matrix of a 2$\times$2 MZI with two PhS and two 3-dB BeS (see Fig. \ref{placeholder}(d)) can be defined as \cite{fang2019design}:
\begin{equation}
    \begin{split}
       &T_{MZI}(\theta, \phi)=U_{BeS}\cdot U_{PhS}(\theta)\cdot U_{BeS}\cdot U_{PhS}(\phi)\\
       &=\begin{pmatrix}
       T_{11} & T_{12} \\
        T_{21} & T_{22}
       \end{pmatrix} =  \begin{pmatrix}
        \frac{e^{i\phi}}{2}(e^{i\theta}-1) & \frac{i}{2}(e^{i\theta}+1) \\
        \frac{ie^{i\phi}}{2}(e^{i\theta}+1) & -\frac{1}{2}(e^{i\theta}-1) 
        \end{pmatrix} 
    \end{split},
    \label{T_MZI}
\end{equation}
where $U_{BeS}$ and $U_{PhS}$ denote the transfer matrices for the 3-dB BeS and the PhS, respectively. Note that the expression for $T_{MZI}$ in \eqref{T_MZI} assumes ideal PhS and BeS. Later, we will augment the MZI transfer matrix to include the effects of uncertainties in phase angles and BeS splitting ratios.\par

\subsection{MZI-based Coherent IPNNs}
A multi-layer perceptron (MLP)-based DNN consists of several consecutive layers of interconnected neurons. Post feature-extraction, the input features ($X_1, \dots, X_{N_1}$) are fed into a series of fully connected layers, followed by a final LogSoftMax activation layer to obtain the probability of each output class ($Y_1, \dots, Y_{N_2}$). Each connection between the neurons is assigned a weight that represents its synaptic plasticity and each neuron is tasked with a multiply-and-accumulate (MAC) operation followed by passing the resultant output through a non-linear activation function ($f_{NAU}$). By introducing non-linearity in the network, the activation functions (e.g., sigmoid, $\tanh$, and Rectified Linear Unit) enable the DNNs to learn complex non-linear relationships \cite{sunny2021survey}. During each training iteration, the weight of each connection in a DNN is incrementally updated to minimize the loss function that quantifies the difference between the expected and the obtained DNN output. \par 

Consider an $N_2\times N_1$ weight matrix $L_m$ representing the edge weights connecting a layer with $N_1$ neurons with a layer with $N_2$ neurons. Using singular value decomposition (SVD) and considering Fig. \ref{placeholder}(c), we have $L_{m}=U\Sigma V^H$, where $U$ and $V$ are unitary matrices with dimensions $N_{2}\times N_{2}$ and $N_{1}\times N_{1}$, respectively. Moreover, $V^H$ denotes the Hermitian transpose of $V_{m}$, and $\Sigma$ is a diagonal matrix consisting of the eigenvalues of $L_{m}$. \par

Reck \textit{et al} \cite{reck1994experimental} first demonstrated that any unitary transformation between optical channels can be realized using a triangular mesh of MZIs. However, Clements \textit{et al} proposed an alternative arrangement of MZIs (see Fig. \ref{placeholder}) to implement unitary transformations with half the physical footprint of the Reck design and a lower optical loss \cite{clements2016optimal}. Therefore, for a given weight matrix $W_{m}=U_{m}\Sigma_{m}V_{m}^H$, this paper assumes the Clements design to represent the unitary matrices $U_{m}$ and $V_{m}^H$. The diagonal matrix $\Sigma_{m}$ can be realized using an array of MZIs to attenuate each channel separately without mixing by terminating one input and one output of each MZI ($\Sigma$ in Fig. \ref{placeholder}). As MZIs can only attenuate optical signals, a global optical amplification is necessary on each output to represent arbitrary diagonal matrices \cite{connelly2007semiconductor}. This scaling factor is realized using the optical gain unit (OGU) $G$ (see Fig. \ref{placeholder}) that includes semiconductor optical amplifiers \cite{haq2020micro}.\par 

\subsection{Imperfections in IPNNs} 
We classify imperfections in IPNNs into three broad categories: (1) uncertainties in phase angles and splitting ratios due to FPVs and thermal crosstalk, (2) non-uniformity in the insertion loss in MZIs due to FPVs, and (3) quantization errors in the phase angles due to low-precision DACs. FPVs in silicon photonic integrated circuits have been studied by comparing the response of identically designed devices on the same die, across multiple dies on the same wafer, across different wafers, and also across different fabrication runs \cite{nikdast2016chip}. In \cite{selvaraja2011soi}, a non-uniformity of up to $\pm$~20.7~nm was observed in the top silicon thickness of dies in a 200~mm SOI wafer. Variations in silicon layer thickness and etch depth have also been shown to result in degraded performance of optical devices, such as photonic switches, microring resonators \cite{beausoleil2011devices}, and MZIs \cite{bogaerts2019layout}. IPNNs are also sensitive to mutual thermal crosstalk. During inferencing in IPNNs, the trained phase settings stored in the digital memory are converted to voltage inputs to the microheater coils using a digital-to-analog converter (DAC) and a voltage driver. The optical phase shift effected by the microheaters is proportional to the square of the applied voltage. Due to such thermal crosstalk among proximal microheaters, the tuned phase shifts may deviate from their expected value \cite{jha2019automated}, thereby leading to incorrect matrix multiplication and faulty inferencing. Modeling the effects of thermal crosstalk requires complex functional simulations and extensive experimental measurements on a taped-out photonic circuit and is beyond the scope of this paper. \par


Optical signal traversing an MZI experiences coupling loss \cite{bahadori2016comprehensive}, absorption loss in the microheaters' metal planes \cite{ding2016broadband}, and propagation loss in the waveguides \cite{bahadori2016comprehensive}. Prior experimental analysis has shown a loss of up to 1.5 dB in a standalone MZI \cite{shokraneh2020theoretical}. This optical loss is non-unform and can vary across MZIs due to FPVs.  Other major sources of imprecisions are the low-precision DACs used to encode the phase angles. Low-precision encoding is especially favored for low-power applications as the energy efficiency of DACs scales exponentially with the number of bits \cite{saberi2011analysis}. This leads to inevitable quantization errors in the phase angles and thereby resulting in incorrect unitary transformations and a degraded inferencing accuracy. We observe that, while all these imperfections affect the MAC operations in coherent photonic neurons, the nature of their impact differs. It is, therefore, necessary to first model them in a standalone manner before simulating their simultaneous impact.   

\textcolor{black}{IPNNs are also susceptible to dynamic errors in the phase shifts due to thermal drifts during programming. Such dynamic errors are expected to be more prevalent in the Clements architecture -- this is because the Clements network does not provide access to the inputs/outputs of individual MZIs and necessitates complex programming schemes with power sampling (e.g., \cite{miller2017setting}). However, such errors have not yet been quantitatively analyzed and, therefore, should be explored in future work.}


\subsection{Related Prior Work}
While the impact of imperfections in silicon photonic devices and optical interconnects have been studied \cite{nikdast2016chip}\cite{mirza2021silicon}, there has barely been any prior work to help understand the cumulative impact of different uncertainties and imprecisions on IPNNs built using imperfect photonic components. Even in the few cases where this has been studied, efforts typically focus on highlighting the impact of one (or in some cases, a select few) source(s) of imperfections. For example, \cite{shokraneh2020diamond} explored the effect of non-uniform insertion loss in the MZIs and uncertainties in the phase angles alone. Both the insertion loss and phase errors are sampled from zero-mean Gaussian distributions, however, the respective distributions are assumed to have the same standard deviation across all the MZIs. The deployment of thermal actuators to compensate for phase errors was proposed in \cite{milanizadeh2019canceling}. However, the micro-heaters in such actuators lead to induced mutual thermal crosstalk among neighboring waveguides. A method to counter the impact of uncertainties using a modified cost function during training and post-fabrication hardware calibration was presented in \cite{ying2020variation}. However, this method only focuses on uncertainties in the phase angles, ignoring the considerable impact of inevitable errors in BeS ($\approx$50\% reduction in network accuracy as we will show later). Moreover, the required hardware calibration necessitates extraction of the effects of FPVs on each MZI in the network using a differential test. But this step becomes increasingly compute-intensive as the network scales up. The modified training also results in up to a 5\% loss in the inferencing accuracy, which is unacceptable for many critical applications (e.g., autonomous driving).\par

In our preliminary work \cite{banerjee2021modeling}, we modeled the impact of uncertainties in the phase angles and splitting ratios in MZIs on the IPNN accuracy and showed that their impact can vary widely based on the position and the tuned phase angles in the affected MZIs. In \cite{banerjee2021optimizing}, we leverage the non-uniqueness of SVD under reflections to propose an optimization technique to improve the IPNN performance in the presence of uncertainties in phase angles while ensuring that the trained weights remain unaffected, thereby guaranteeing no loss in the nominal inferencing accuracy. This paper extends our analysis in \cite{banerjee2021modeling,banerjee2021optimizing} by considering, for the first time, the impact of spatially correlated uncertainties in PhS and BeS, non-uniform MZI insertion loss, and quantization error in PhS. \par  

This paper presents a hierarchical and comprehensive analysis of the impact of various uncertainties and imprecisions that can affect IPNNs. These imperfections can stem from various sources, e.g., FPVs, run-time thermal crosstalk, and low-precision DACs used for phase encoding. Fig. \ref{bottomup} shows an overview of the proposed bottom-up approach using which we model each of these imperfections. We also present a case study of an IPNN under multiple simultaneous imperfections and demonstrate how its inferencing accuracy is affected under several different realistic scenarios.  

%% file: random.tex
In this section, we systematically analyze the impact of uncertainties in the phase angle and splitting ratio in MZIs on the inferencing accuracy of IPNNs in a bottom-up approach. 


\subsection{Component-Level: Phase Shifters and Beam Splitters}
\begin{figure}[t]
  \centering
  \includegraphics[width=0.48\textwidth]{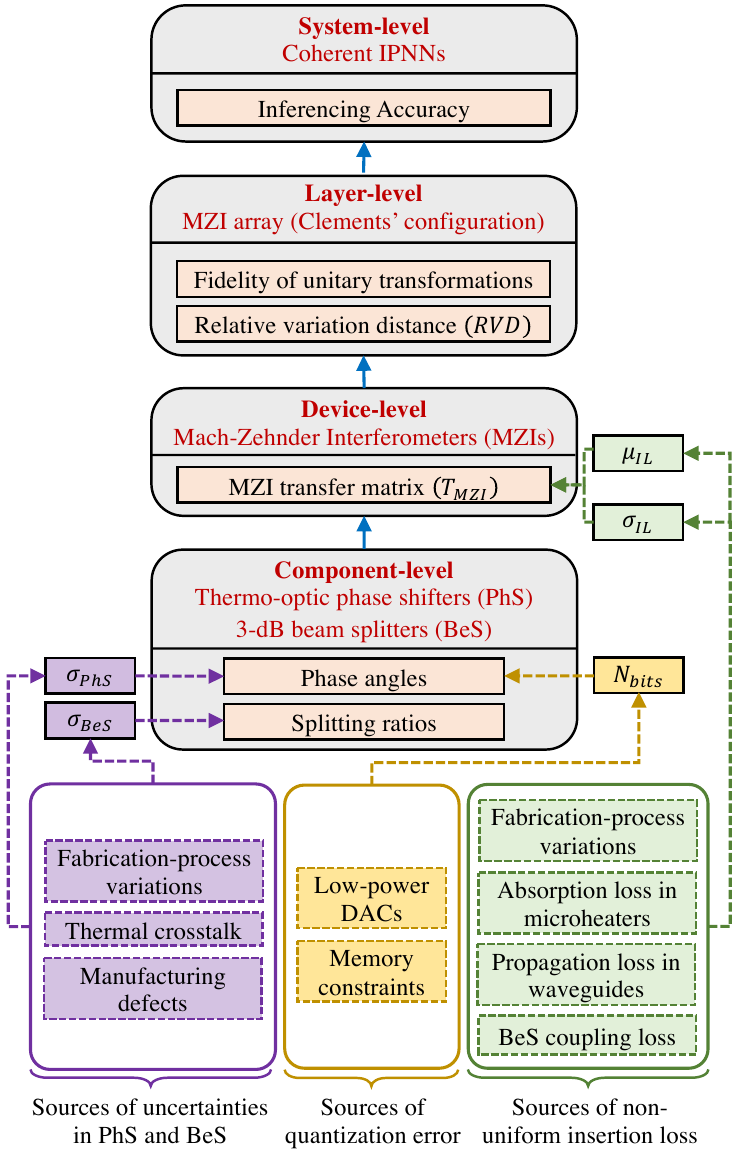}
  \caption{Proposed bottom-up modeling framework.}
  \vspace{-1em}
  \label{bottomup}
\end{figure}
We consider using thermo-optic PhS in IPNNs, which have been widely employed in coherent IPNNs \cite{shokraneh2020diamond}\cite{shokraneh2020theoretical}. The temperature-dependent phase change ($\Phi$) in a thermo-optic phase shifter can be modeled as \cite{jacques2019optimization}: 
\begin{equation}
    \Phi=\left(\frac{2\pi l}{\lambda_0}\right)\cdot \left(\frac{dn}{dT}\right)\cdot \Delta T.
    \label{del_phi}
\end{equation}
Here, $l$ is the length of the thermo-optic phase shifter and $\lambda_0$ is the optical wavelength. Also, $\frac{dn}{dT}\approx$ 1.8$\times10^{-4} K^{-1}$ is the thermo-optic coefficient of silicon at $\lambda_{0}=$~1550~nm and temperature $T=$~300~$K$ \cite{walls2007quantum}, and $\Delta T$ is the temperature change. \textcolor{black}{Under FPVs, the length $l$ of the phase shifter can change. This, in turn, results in a deviated phase angle. From (2), observe that with all other parameters remaining constant and for a small shift, $\Delta l$, in the length, $\Delta\Phi/\Phi=\Delta l/l$. Therefore, an $x\%$ shift in $l$ leads to an $x\%$ shift in $\Phi$.}
During \textit{in-situ} training of IPNNs, the phase angles in PhS are applied using thermal actuators (i.e., microheaters). \textcolor{black}{Let $\Delta T$ be the change in temperature and $P$ be the corresponding heater power consumption required for a phase shift $\Phi$. From \cite{jacques2019optimization}, we then have $\Delta T=\frac{P}{(G\cdot A)}$. Here, $G$ denotes the thermal conductance between the heated waveguide and the heat sink and $A$ denotes the cross-sectional area traversed by the heat flow. If the heater coil has a resistance $R$, the voltage from the DC source is given by $\Delta T=\frac{V^{2}}{(G\cdot A\cdot R)}$. Leveraging (\ref{del_phi}), we therefore have: 
\begin{equation}
    \Phi=\left(\frac{2\pi l}{\lambda_0}\right)\cdot \left(\frac{dn}{dT}\right)\cdot \frac{V^2}{G\cdot A\cdot R}=K\cdot V^2,
    \label{precision_del_phi}
\end{equation}
where $K$ is a constant of proportionality that depends on the structure of the phase shifter. Changes in the supply voltage $V$ (due to voltage droop and IR loss) lead to a deviated $\Phi$. From (\ref{precision_del_phi}), we have, for a small voltage shift $\Delta V$, $\Delta\Phi/\Phi=2\cdot(\Delta V/V)$. Therefore, an $x\%$ shift in $V$ results in a 2$x\%$ shift in $\Phi$.} Additionally, mutual thermal crosstalk among neighboring actuated waveguides, which are placed in proximity in IPNNs, imposes phase errors in $\phi$ and $\theta$ (see  Fig.~\ref{placeholder}(d)). While methods such as \cite{milanizadeh2019canceling} have been proposed to mitigate thermal crosstalk during the transient state (training), prior work does not address random perturbations in the phase angles during steady-state inferencing. Due to perturbations in the phase angles ($\theta$ and $\phi$ in (\ref{T_MZI})), $T_{MZI}$ will deviate from its intended form, resulting in faulty matrix multiplication. 

Considering the classical, lossless 2$\times$2 beam-splitter schematic shown in Fig. \ref{placeholder}(d), the electric fields at the output $\tilde{E}_{0/1}$ can be attributed to the transmitted electric-field $E_0$ and the reflected electric-field $E_1$ based on \cite{fang2019design}:
\begin{equation}
    \begin{pmatrix}
        \tilde{E_0} \\
        \tilde{E_1} 
        \end{pmatrix}
        = \begin{pmatrix}
        r_{00} & it_{10} \\
        it_{01} & r_{11} 
        \end{pmatrix}
        \begin{pmatrix}
        E_0 \\
        E_1
        \end{pmatrix}.
        \label{E01}
\end{equation}
Here, $r$ and $t$ denote the reflectance and transmittance associated with each path, respectively. Note that $r_{00}^2+t_{01}^2=$~1 and $r_{11}^2+t_{10}^2=$~1. For symmetric BeS, $r_{00}=r_{11}=r$ and $t_{01}=t_{10}=t$. Additionally, for ideal 3-dB BeS, $r=t=\frac{1}{\sqrt{2}}$. However, under uncertainties, $r$ and $t$ will deviate from $\frac{1}{\sqrt{2}}$; this results in unbalanced and imperfect BeS \cite{liu2016compensation, nikdast2016chip}. Unlike PhS, BeS are passive devices and once fabricated, we cannot actively change their $r$ and $t$ values during IPNN training (e.g., in a passive directional coupler).\par

Prior studies have shown an error of 0.21 radian in the tuned phase angles in PhS for mature fabrication processes \cite{flamini2017benchmarking}. This corresponds to $\frac{0.21}{2\pi}\times 100$ $\approx$~3.3\% of the maximum possible phase angle, $2\pi$. Taking this into consideration, we perturb $\theta$ and $\phi$ using a Gaussian distribution with mean ($\mu$) set to their nominal tuned values (obtained from training) and multiple values of standard deviation in the range \mbox{$0.005\cdot 2\pi\leq\sigma\leq 0.05\cdot 2\pi$}. Although a deviation of only 3.3\% of $2\pi$ is expected in mature fabrication processes, we consider this wider range to demonstrate IPNN accuracy for emerging immature processes. While a deviation of 1--2\% is typically expected in the $r$ and $t$ parameters in BeS \cite{flamini2017benchmarking}, we vary them using a similar distribution as PhS---Gaussian with $\mu=\frac{1}{\sqrt{2}}$ and $0.005\cdot \frac{1}{\sqrt{2}}\leq\sigma\leq 0.05\cdot \frac{1}{\sqrt{2}}$---for a fair comparison of their impact. In the rest of the paper, $\sigma_{PhS}$ refers to $\frac{\sigma}{2\pi}$ for PhS, and $\sigma_{BeS}$ refers to $\sqrt{2}\sigma$ for BeS.

\subsection{Device-Level: MZIs}
Variations in $\theta$ ($\Delta\theta$) and $\phi$ ($\Delta \phi$) phase angles in PhS can result in deviations in the MZI transfer matrix ($T_{MZI}$) defined in (\ref{T_MZI}). Such deviations can be modeled as:
\begin{equation}
\begin{split}
       \Delta T_{MZI}(\theta, \phi) &= \frac{\partial T_{MZI}(\theta, \phi)}{\partial \theta}\Delta \theta + \frac{\partial T_{MZI}(\theta, \phi)}{\partial \phi}\Delta \phi \\
    & =\begin{pmatrix}
    \frac{ie^{i(\phi+\theta)}}{2} & -\frac{e^{i\theta}}{2} \\
    -\frac{e^{i(\phi+\theta)}}{2} & -\frac{ie^{i\theta}}{2}
    \end{pmatrix}
    \Delta\theta \\&+ 
    \begin{pmatrix}
    \frac{ie^{i\phi}}{2}(e^{i\theta}-1) & 0 \\
    -\frac{e^{i\phi}}{2}(e^{i\theta}+1) & 0
    \end{pmatrix}
    \Delta\phi.  
\end{split}
\label{del_T_MZI}
\end{equation}
Let the relative changes in $\theta$ and $\phi$ be $K_{\theta}=\frac{\Delta\theta}{\theta}$~and \mbox{$K_{\phi}=\frac{\Delta\phi}{\phi}$}, respectively. We assume $K_{\theta}=K_{\phi}=K$ as the two PhS, corresponding to $\theta$ and $\phi$, are in proximity (see Fig. \ref{placeholder}(d)). This assumption is made to simplify the analyses only in this subsection. In all subsequent analyses independent variations are considered in $\theta$ and $\phi$. Thus, from (\ref{del_T_MZI}), we have: 
\begin{equation}
    \Delta T_{MZI}(\theta, \phi)=K
    \begin{pmatrix}
    (\theta+\phi)\frac{ie^{i(\theta+\phi)}}{2}-\phi\frac{ie^{i\phi}}{2} & -\theta\frac{e^{i\theta}}{2} \\
    -(\theta+\phi)\frac{e^{i(\theta+\phi)}}{2}-\phi\frac{e^{i\phi}}{2} & -\theta\frac{ie^{i\theta}}{2}
    \end{pmatrix}.
    \label{del_T_MZI_2}
\end{equation}

\begin{figure}[t]
\centering
\subfigure[$T_{11}$]{
\includegraphics[scale=0.2]{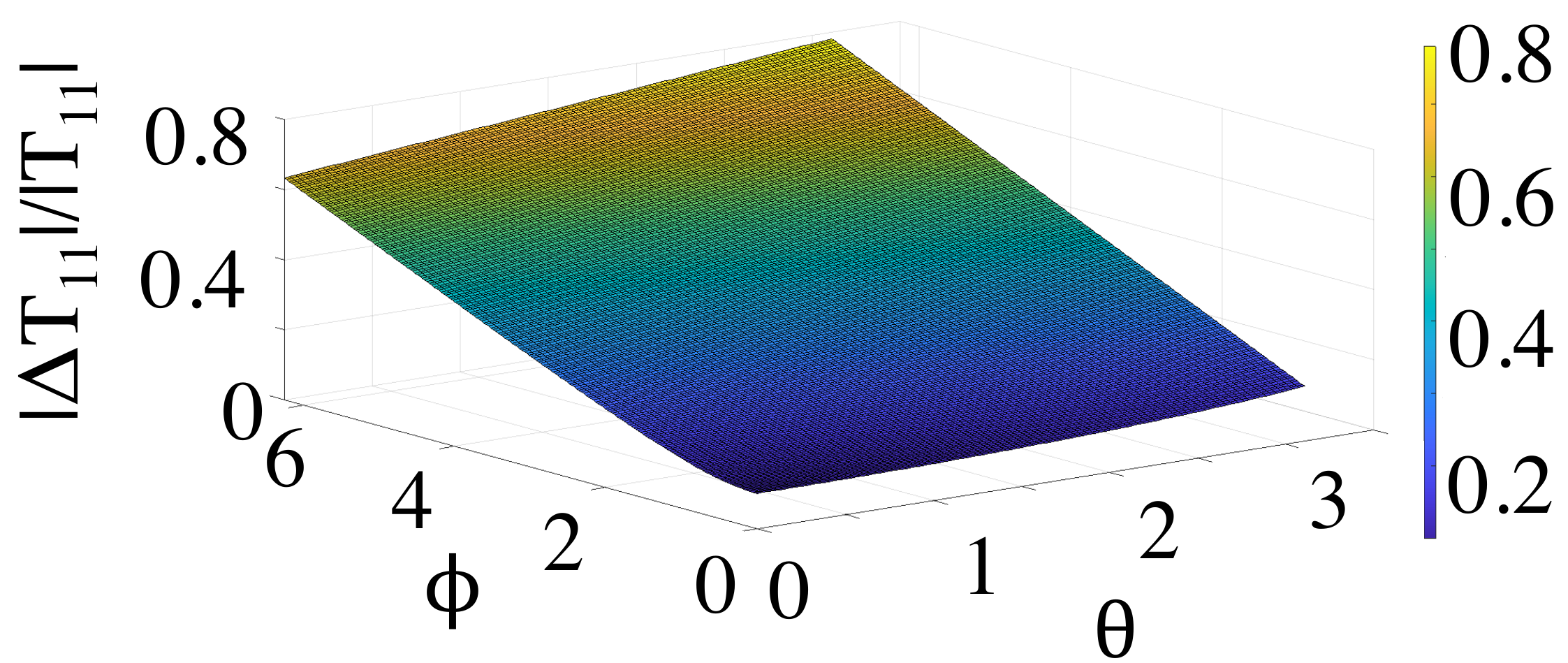}\label{fab}}\hspace{-0.3em}%
\subfigure[$T_{12}$]{
\includegraphics[scale=0.2]{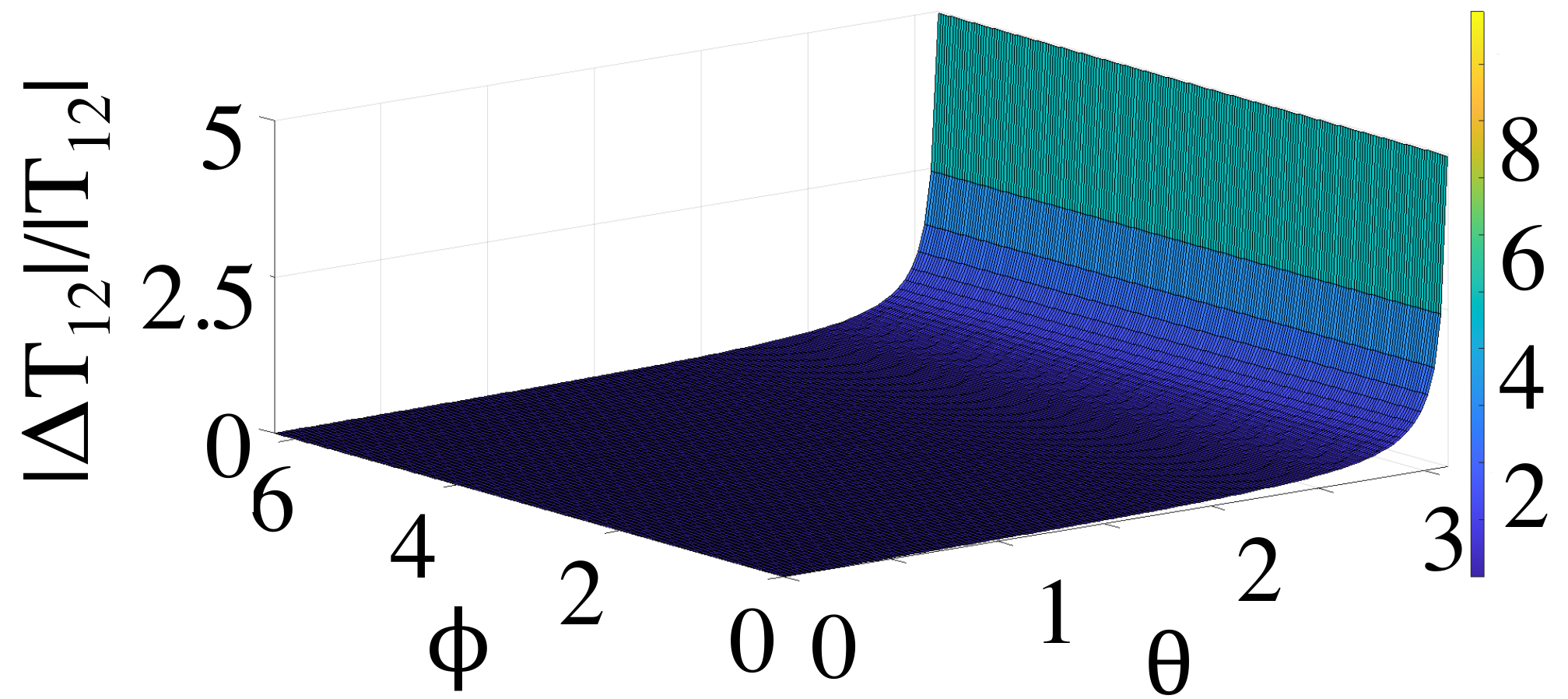}\label{slps}}%
\newline
\subfigure[$T_{21}$]{
\includegraphics[scale=0.2]{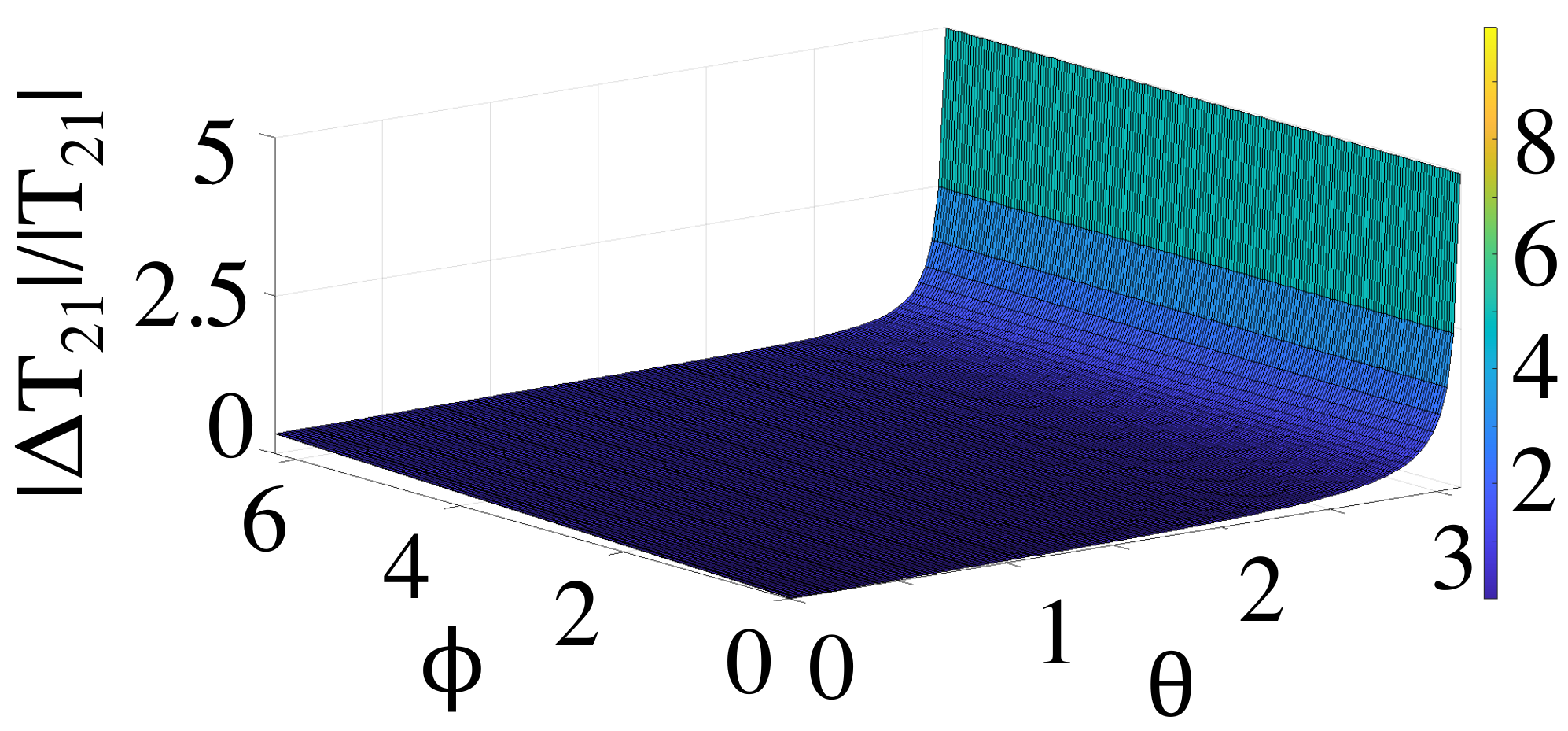}\label{total_shift}}\hspace{-0.3em}%
\subfigure[$T_{22}$]{
\includegraphics[scale=0.2]{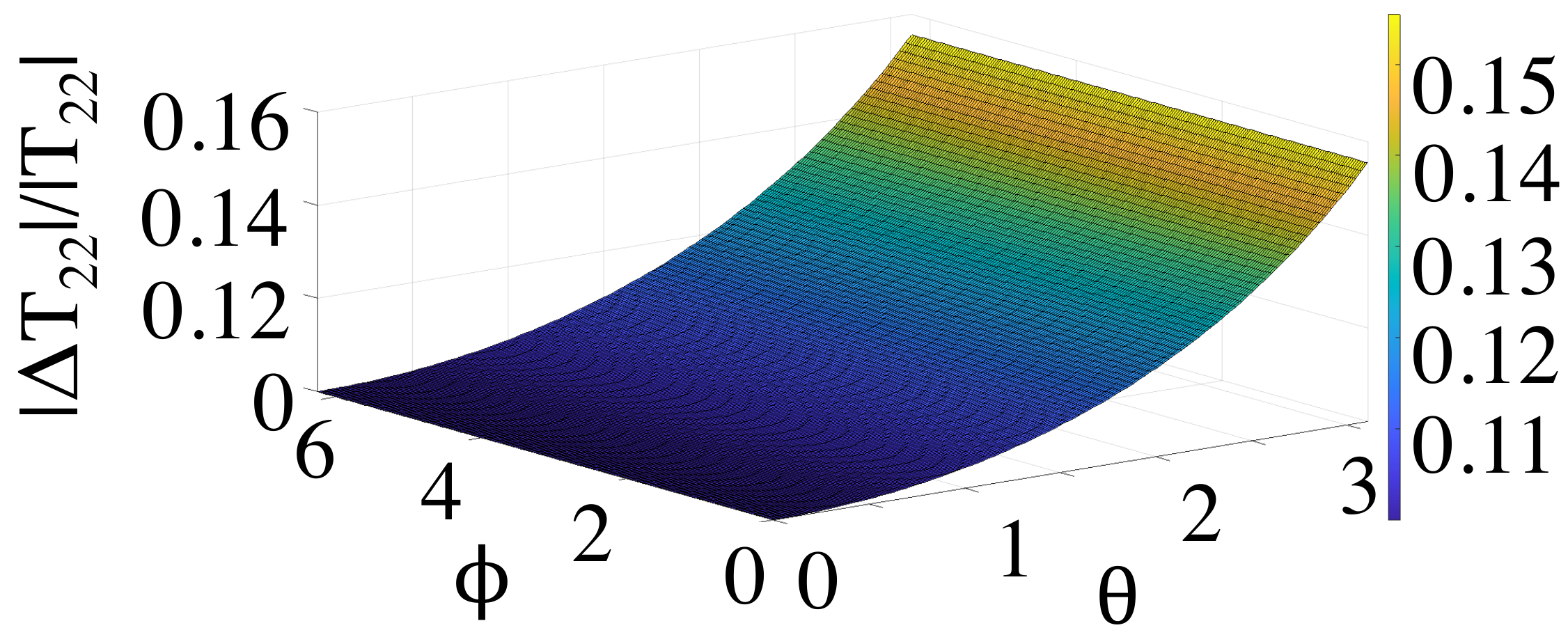}\label{fab}}%
\vspace{-0.05in}
\caption{Magnitude of variation in the absolute value of elements in $T_{MZI}$ (see (1)) relative to the modulus of their nominal values.}\label{Tmn_results}
\vspace{-0.2in}
\end{figure}

Using (\ref{T_MZI}) and (\ref{del_T_MZI_2}), Fig. \ref{Tmn_results} shows the magnitude of deviation for each of the four elements in $T_{MZI}$ (i.e., $T_{11}$, $T_{12}$, $T_{21}$, and $T_{22}$) relative to the modulus of their nominal values for different values of $\theta$ and $\phi$ with $K=$~0.05, chosen as an example. We find that the relative deviation increases monotonically as $\theta$ and $\phi$ increase. This indicates that the transfer matrix of an MZI with higher tuned phase angles is more sensitive to phase errors. This observation holds for all $K$ at the standalone device-level. However, we will later show that when MZIs are connected in an array, the accuracy of the resulting unitary transformation depends on both the tuned phase angles and the position of the affected MZIs. \par

The proposed $T_{MZI}$ model in (\ref{T_MZI}) assumes ideal 3-dB BeS with $r_{00}=r_{11}=t_{01}=t_{10}=\frac{1}{\sqrt{2}}$. However, under uncertainties in BeS, this model changes to:
\begin{equation}
    T_{MZI}(\theta, \phi)=\begin{pmatrix}
    rr'e^{i(\theta+\phi)}-tt'e^{i\phi} & ir'te^{i\theta}+it'r \\
    it're^{i(\theta+\phi)}+itr'e^{i\phi} & -tt'e^{i\theta}+rr'
    \end{pmatrix},
    \label{T_MZI_general}
\end{equation}
where $r$ and $r'$ denote the reflectances of the first and the second beam splitter, respectively, while $t$ and $t'$ denote their transmittances, with each of these deviated from $1/\sqrt{2}$. We replace the ideal MZI transfer matrix (in (1)) with its general form (in (6)) for the rest of our analysis.

\subsection{Layer-Level: MZI Array}
Under uncertainties in PhS and BeS, $T_{MZI}$ deviates, and consequently, the matrix represented by the array can vary from the intended unitary matrix. We use the relative-variation distance (RVD) as a figure-of-merit to quantify the difference between the intended unitary matrix ($\tilde{U}$) and the deviated unitary matrix ($U$). This is given by: 
\begin{equation}
    RVD(U, \tilde{U})=\sum\limits_{m}\sum\limits_{n}\frac{\left|U_{m,n}-\tilde{U}_{m,n}\right|}{\left|\tilde{U}_{m,n}\right|},
    \label{RVD}
\end{equation}
where $U_{m,n}$ and $\tilde{U}_{m,n}$ denote the element at the $m$\textsuperscript{th} row and $n$\textsuperscript{th} column of unitary matrix $U$ and $\tilde{U}$, respectively.\par

\begin{figure}[t]
  \centering
  \includegraphics[scale=0.42]{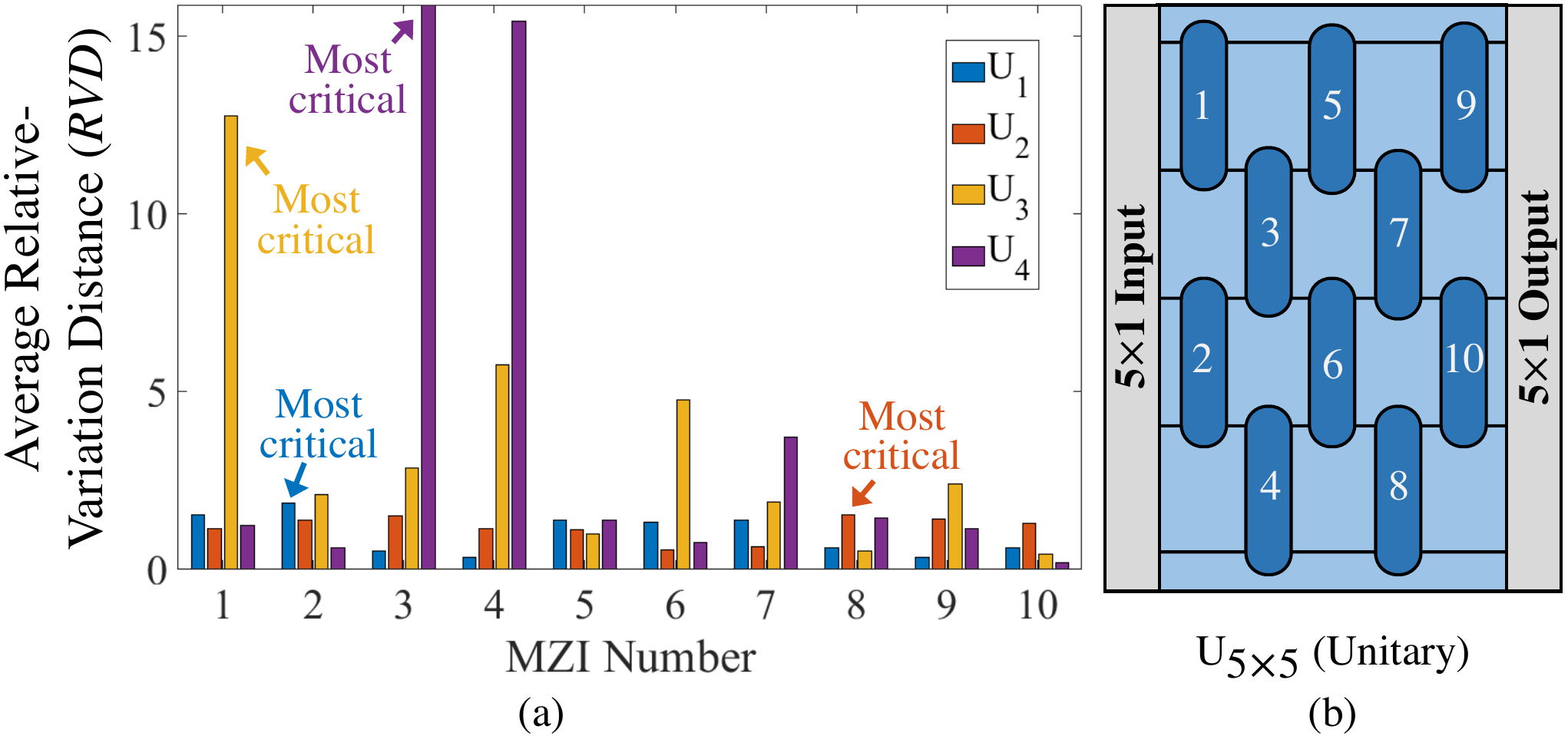}
  \caption{(a) Average $RVD$ for four random 5$\times$5 unitary matrices with one MZI under variations at a time. (b) An MZI array (including the MZI numbers) to represent any 5$\times$5 unitary matrix (see Fig. \ref{placeholder}).}
  \vspace{-1em}
  \label{U1234}
\end{figure}

Different elements of a unitary transfer matrix are affected by different subsets of MZIs in the array. Therefore, variations in each MZI will have varying impacts on the overall $RVD$ defined in \eqref{RVD}. This is indeed the case as is shown in Fig. \ref{U1234}. We consider four randomly generated 5$\times$5 unitary matrices with uncertainties in the PhS and BeS. For each matrix, we introduce variations in one MZI at a time. For each MZI, we perform 1000 Monte Carlo iterations and calculate the average $RVD$. In each iteration, the MZI parameters ($\theta$,~$\phi$,~$r$,~$r'$,~$t$,~$t'$) corresponding to the faulty MZI are chosen from a Gaussian distribution with $\sigma_{PhS}=\sigma_{BeS}=$~0.05. From Fig. \ref{U1234} we observe that there is a significant variation in the average $RVD$ corresponding to different MZIs representing the same unitary matrix. Note also that the distribution of average $RVD$ across the MZIs differs across the four example unitary matrices. Clearly, the impact of uncertainties in an MZI on the accuracy of the unitary transformation varies based on both the tuned phase angles in the MZI and its position in the array. As a result, the resilience of different IPNNs against uncertainties may also vary even though they are structurally mapped to the same MZI array. Therefore, any study on the reliability of IPNNs must be performed post-training in order to consider the tuned phase angles. \par

\subsection{System-Level: Coherent IPNNs}
\begin{figure}[t]
  \centering
  \includegraphics[width=0.45\textwidth]{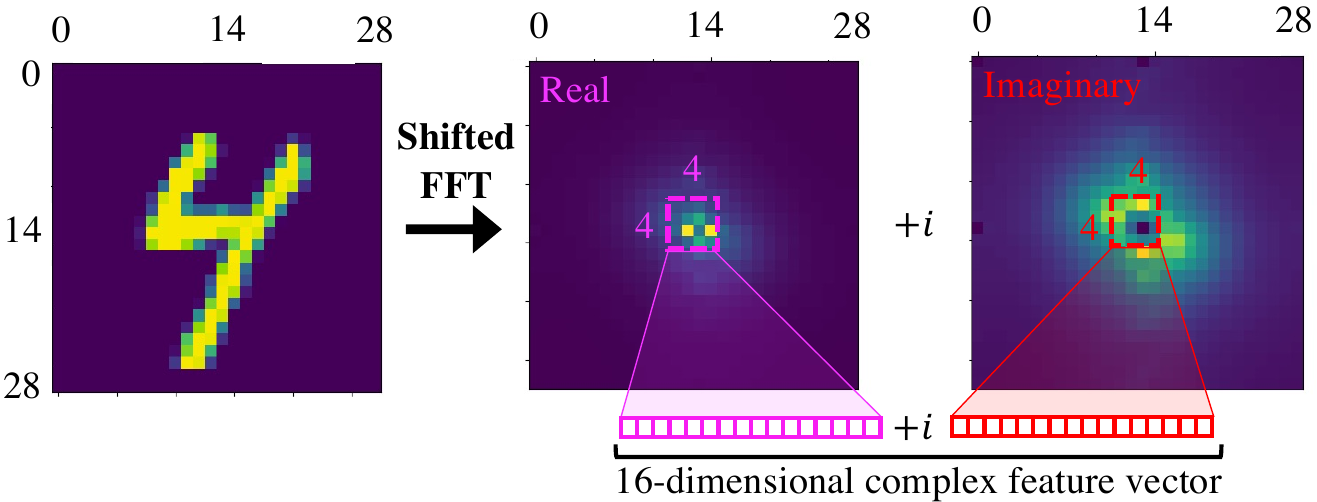}
  \caption{Illustration of the 16-dimensional complex feature vector extraction from an MNIST image using shifted FFT.}
  \vspace{-1em}
  \label{shiftFFT}
\end{figure}

Uncertainties in the MZI parameters lead to faulty matrix multiplication in the linear layers, hence imposing inferencing accuracy loss in IPNNs. To better understand the impact of such uncertainties, we present a case study of an IPNN handling the MNIST hand-written digit classification task. To convert the 28$\times$28~$=$~784 dimensional real-valued images in the MNIST dataset to complex-valued vectors, we consider the shifted fast Fourier transform (FFT) of each image. This results in a 784-dimensional complex-valued vector for each image (see Fig. \ref{shiftFFT}). To compress the feature vector, we consider the values within the 4$\times$4 region at the center of the frequency spectrum. Compared to the baseline accuracy of 94.12\% with the 28$\times$28 feature vector, the 4$\times$4 case results in a negligible 0.26\% accuracy loss. \par

In the IPNN architecture considered in the case study, fully connected feedforward networks with one input layer and two hidden layers of 16-complex valued neurons are implemented using the Clements design \cite{clements2016optimal}. Each linear layer is followed by the nonlinear Softplus function applied to the modulus of the complex numbers. To model intensity measurement, a modulus squared nonlinearity is applied after the output layer. This is followed by a final LogSoftMax layer to obtain a probability distribution. We use a cross-entropy loss function during training \cite{cover2006elements}. \par

We implement the three weight matrices corresponding to the neurons in the input and the two hidden layers in the IPNN using MZI arrays. Based on the network architecture, the dimensions of the weight matrices are 16$\times$16 (input layer, L0), 16$\times$16 (first hidden layer, L1), and 10$\times$16 (second hidden layer, L2). Note that uncertainties in PhS and BeS in IPNNs may have different distributions and correlations. Therefore, we design three experiments ($EXPs$) to analyze the impact of uncertainties in the MZI arrays on the IPNN inferencing accuracy under different uncertainty scenarios:

\begin{itemize}
    \item $EXP_1$: Uncertainties from the same distribution are present across all the MZIs in an IPNN.
    \item $EXP_2$: Uncertainties from the same distribution are present across most MZIs, with localized uncertainties of higher magnitude affecting a few proximal MZIs. Such scenarios may arise as a result of a thermal hotspot in the network or due to manufacturing defects.
    \item $EXP_3$: Spatially correlated uncertainties where proximal MZIs encounter correlated magnitude of uncertainties. Such scenarios have often been observed in experimental studies on FPVs in integrated photonic circuits \cite{lu2017performance}. 
\end{itemize}

\subsubsection{$EXP_1$: Global Uncertainties}
\begin{figure}[t]
  \centering
  \includegraphics[scale=0.49]{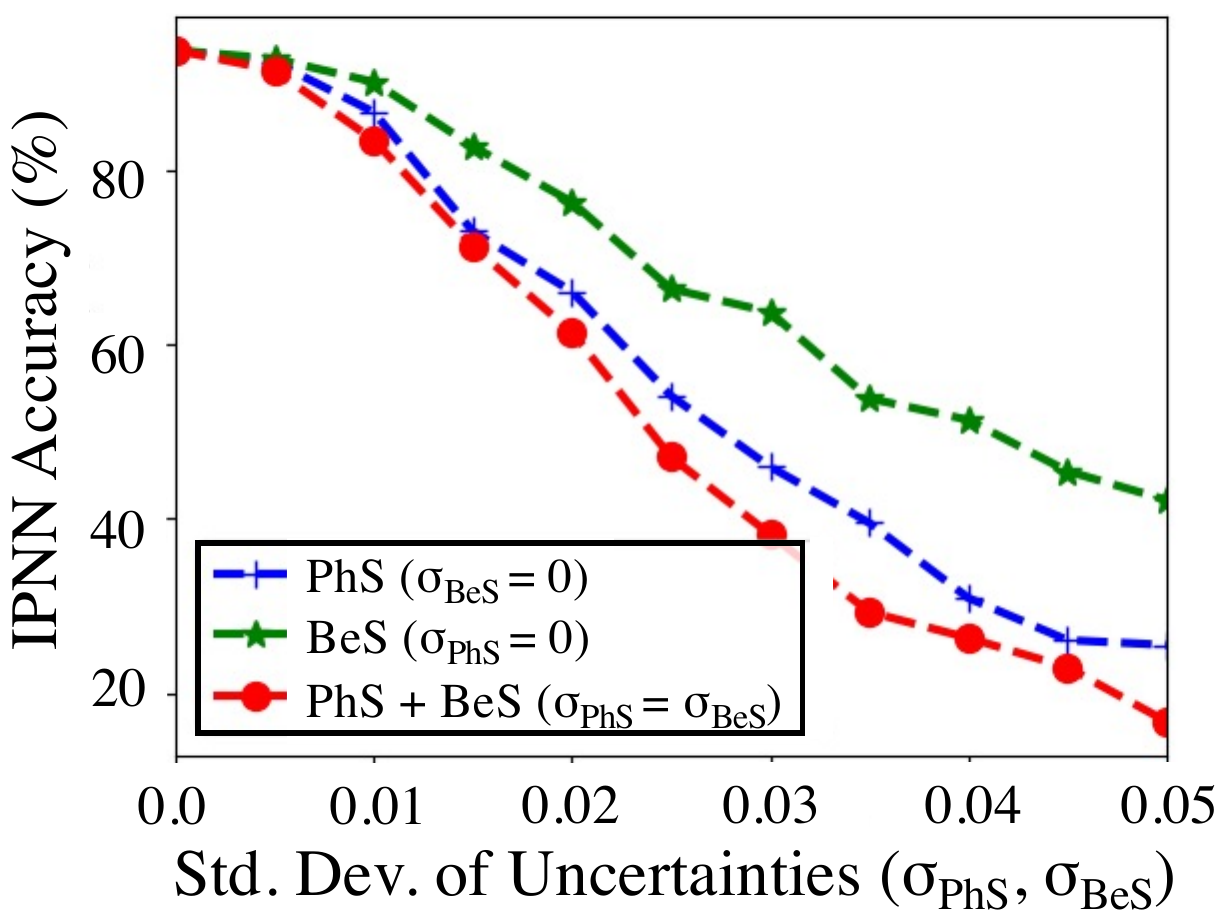}
  \caption{Impact of random uncertainties in the IPNN components (PhS and BeS) on the IPNN accuracy ($EXP_1$).}
  \vspace{-1em}
  \label{compress_EXP1}
\end{figure}

In this experiment, we simulate the uncertainty scenario in which all the MZIs have random uncertainties sampled from the same Gaussian distribution; uncertainties are considered in PhS only, BeS only, and both PhS and BeS. For the PhS-only case, we select several $\sigma_{PhS}$ (see Section III.A) and consider $\sigma_{BeS}=$~0. For each $\sigma_{PhS}$, we perform 1000 Monte Carlo iterations. In each iteration, we calculate the inferencing accuracy using the 10,000 test images in the MNIST dataset. The use of 1000 Monte Carlo iterations is formally justified based on the fact that with a 95\% confidence interval, the maximum margin of error in the mean of the inferencing accuracy is 6.27\%, which is within the acceptable range \cite{acceptablemargin}. The same steps are repeated for the BeS-only case ($\sigma_{PhS}=$~0), and for the both PhS and BeS case ($\sigma_{PhS}=\sigma_{BeS}$), the results of which are shown in Fig. \ref{compress_EXP1}. For all these cases, the accuracy declines steeply as $\sigma_{PhS, BeS}$ increases; this indicates the high sensitivity of the IPNNs to random uncertainties. As expected, the inferencing accuracy drops the most when uncertainties are present in both PhS and BeS. We also observe that uncertainties in PhS have a higher impact on IPNN accuracy compared to those in BeS. Therefore, techniques to improve IPNN robustness should prioritize countering uncertainties in PhS over those in BeS. It helps that, contrary to the static BeS, PhS are tunable. In prior work \cite{banerjee2021optimizing}\cite{banerjee2021champ}\cite{banerjee2021pruning}, we have shown that IPNNs can indeed be made more robust by optimizing the phase angles.  \par

\subsubsection{$EXP_2$: Global Uncertainties with Regional Perturbations}
\begin{figure}[t]
  \centering
  \includegraphics[width=0.48\textwidth]{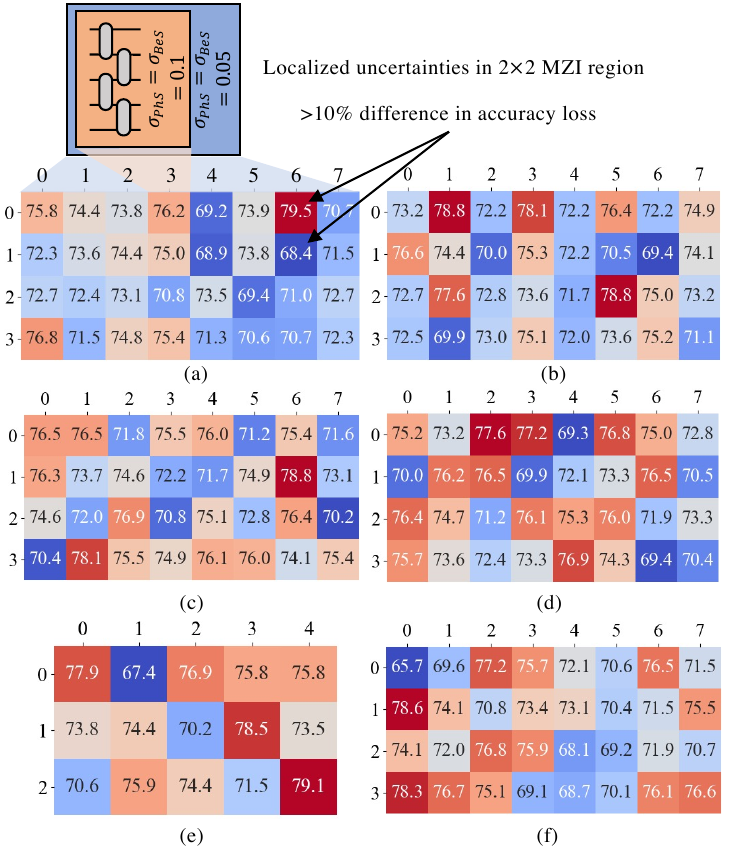}
  \caption{Accuracy loss (\%) due to regional perturbations in IPNN linear layers ($EXP_2$): (a) $U_{L0}$, (b) $V^H_{L0}$, (c) $U_{L1}$, (d) $V^H_{L1}$, (e) $U_{L2}$, and (f) $V^H_{L2}$.}
  \vspace{-1.7em}
  \label{heatmap}
\end{figure}

To find the impact of localized uncertainties on the IPNN accuracy, we divide the IPNN into different regions, each consisting of four MZIs (considered as an example) arranged in a 2$\times$2 grid. We insert random perturbations with $\sigma_{PhS}=\sigma_{BeS}=$~0.1 in a selected region while the remaining regions have uncertainties with $\sigma_{PhS}=\sigma_{BeS}=$~0.05. Note that we consider a high $\sigma$ for the localized uncertainties to simulate the impact of catastrophic manufacturing defects (e.g., microheaters permanently turned off, thermal hotspots, and lattice defects in SOI waveguides\cite{foster2006optical}). For each selected region, we again consider 1000 Monte Carlo iterations (similar to $EXP_1$) and calculate the reduction in the mean inferencing accuracy from the nominal case. \par

The three linear layers in our IPNN (L0, L1, and L2) can be represented by six unitary multipliers. The impact of regional perturbations in these unitary multipliers on the classification accuracy (experiment $EXP_2$) is presented as heatmaps in Fig.~\ref{heatmap}. Figs. \ref{heatmap}(a)--(b) correspond to the $U$ and $V^H$ matrices of L0 while Figs.~\ref{heatmap}(c)--(d) and Figs.~\ref{heatmap}(e)--(f) correspond to L1 and L2, respectively. Note that for all these cases, the diagonal matrix $\Sigma$ is assumed to be error-free with the singular values arranged in random order. Each box in the heatmaps corresponds to a region (i.e., a 2$\times$2 MZI grid with four MZIs) with the height (width) of the layer increasing vertically (horizontally). The value (color) in each box signifies the accuracy loss when a regional perturbation is applied to the corresponding region. From experiment $EXP_1$ (Fig. \ref{compress_EXP1}), we know that the reduction in IPNN accuracy under a global uncertainty of $\sigma_{PhS}=\sigma_{BeS}=$~0.05 is 74.98\%. Fig.~\ref{heatmap} shows that even under regional perturbations, the accuracy loss hovers around 74.98\%. However, in some regions, the regional perturbations result in a decreased accuracy loss (e.g., the region in row 2 column 5 in Fig. \ref{heatmap}(a)), whereas in others they exacerbate the impact of global uncertainties (e.g., the region in row 3 column 0 in Fig.~\ref{heatmap}(f)). In fact, Fig. \ref{heatmap}(a) shows an example where accuracy losses due to regional perturbations in two contiguous regions differ by more than 10\%. Moreover, note that the low- and high-impact regions are arranged randomly in each unitary multiplier. This shows that the impact of localized uncertainties in MZIs can differ significantly and some MZIs are more critical than others (see also Fig. \ref{U1234}). Such an analysis of the impact of localized uncertainties can, therefore, help IPNN designers to develop reliability measures targeted toward MZIs in the high-impact regions in the network. \par

\subsubsection{$EXP_3$: Spatially Correlated Uncertainties}
Uncertainties in phase angles and splitting ratios originating from fabrication-process variations and thermal crosstalk can be spatially correlated. Experimental studies have shown that proximal devices on a wafer encounter, to some degree, similar variations in waveguide thickness and width \cite{mirza2021silicon}. Similarly, thermal crosstalk is also expected to be localized where a hotspot formed on a chip can affect neighboring devices. However, considering such spatially-correlated uncertainties has proven to be challenging for silicon photonic integrated circuits \cite{chrostowski2016schematic}. The Monte-Carlo approach where we apply the same random variation to all the IPNN components ($EXP_1$) solely captures the common-mode variability while ignoring spatially correlated uncertainties. This is extended in $EXP_2$ to allow for differential variations within a 2$\times$2 region in an MZI array. This approach is appropriate for electrical devices that are significantly smaller than operation wavelengths. Therefore, differential uncertainties need to be considered only for a few components that require matching (e.g., matching resistor pairs in differential amplifiers) \cite{sedra1998microelectronic}. As photonic devices are much larger than the operation wavelength, small changes in waveguide dimensions can result in large uncertainties in the device performance. As a result, uncertainties in all (and not just in a few) components in photonic integrated circuits are expected to be spatially correlated \cite{lu2017performance}.\par

\begin{figure}[t]
  \centering
  \includegraphics[width=0.48\textwidth]{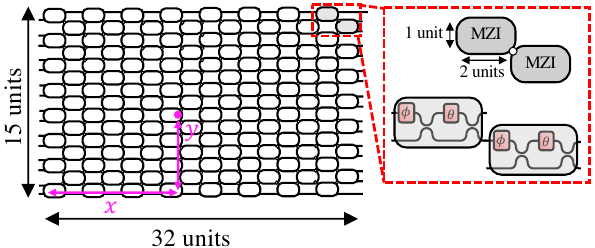}
  \caption{MZIs connected in an array to represent a 16$\times$16 unitary multiplier. A typical MZI has an aspect ratio of $\approx$~2:1 \cite{lagali2000generalized}. Therefore, the MZIs in this multiplier can be represented in a 15 units~$\times$~32 units grid as shown. Each MZI is assumed to be 2 units long and 1 unit wide. In our analysis, one unit corresponds to $l_{MZI}/2$, where $l_{MZI}$ is the length of an MZI. Note that while we consider MZIs with a 2:1 aspect ratio, our approach is applicable for any MZI dimensions.}
  \vspace{-1.7em}
  \label{fig6}
\end{figure}

 Prior work (e.g., \cite{lu2017performance}) has shown that accurately capturing correlated uncertainties requires $O\left(N^2\right)$ correlation parameters for $N$ components, and determining these parameters from a physical layout has proven to be challenging. We address this by proposing an approach that simulates the impact of such correlated uncertainties in IPNNs by defining the correlation parameters (in our case, correlation length) in terms of the MZI dimensions. As a result, this approach is agnostic to optical lithography and can be extended to all emerging IPNNs. \par
 
\begin{figure*}[t]
\centering

\subfigure[Uncorrelated (non-radial)]{
\includegraphics[width=.245\textwidth]{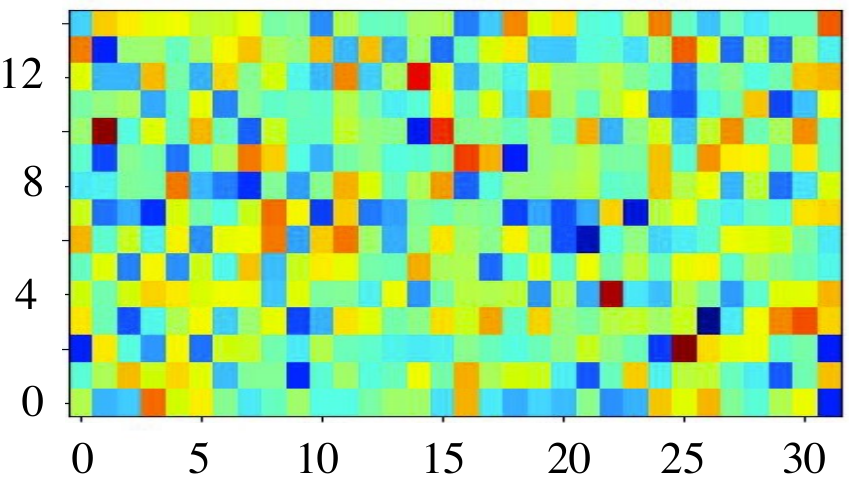}
}
\subfigure[Uncorrelated (radial)]{
\includegraphics[width=.274\textwidth]{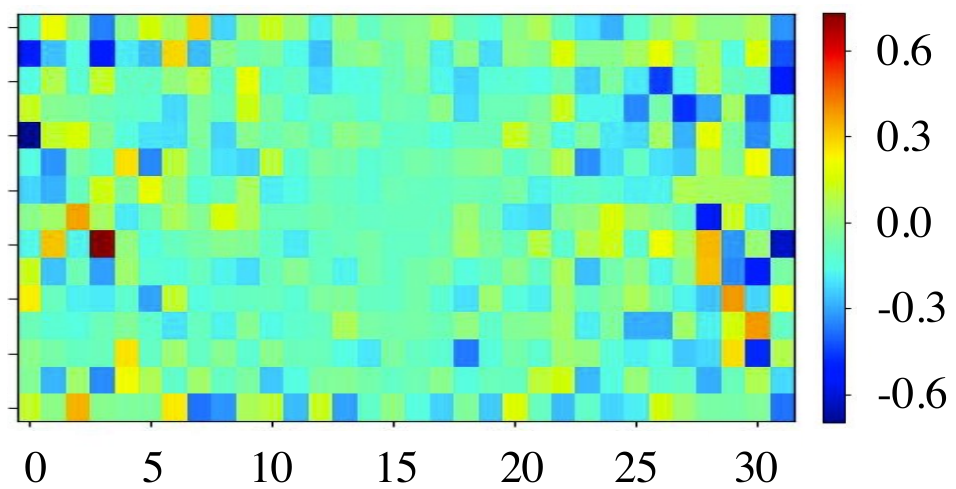}
}
\subfigure[Gaussian kernel ($L=$~4)]{
\includegraphics[width=.275\textwidth]{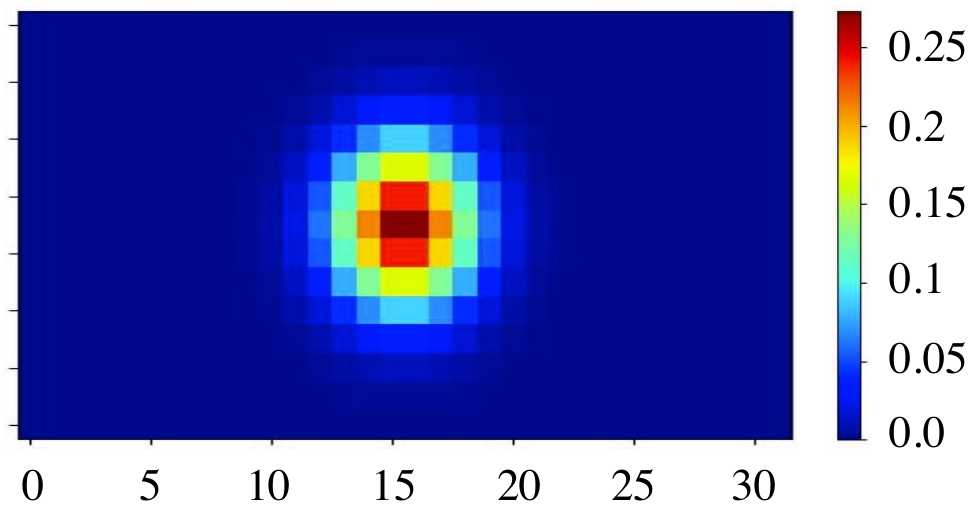}
}
\newline
\subfigure[Correlated (non-radial, $L=$~2)]{
\includegraphics[width=.23\textwidth]{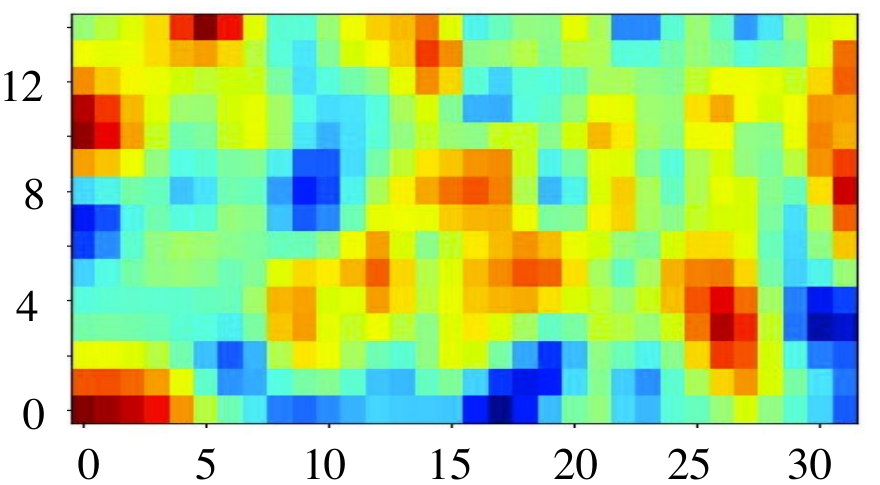}
}
\subfigure[Correlated (non-radial, $L=$~4)]{
\includegraphics[width=.21\textwidth]{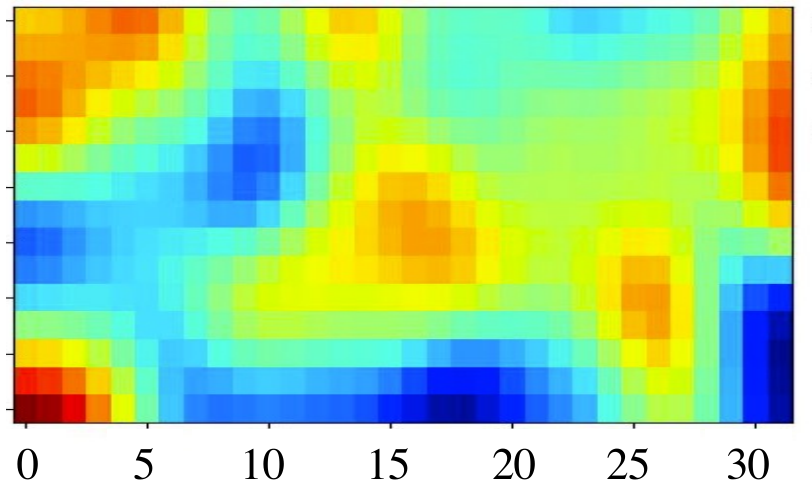}
}
\subfigure[Correlated (non-radial, $L=$~8)]{
\includegraphics[width=.21\textwidth]{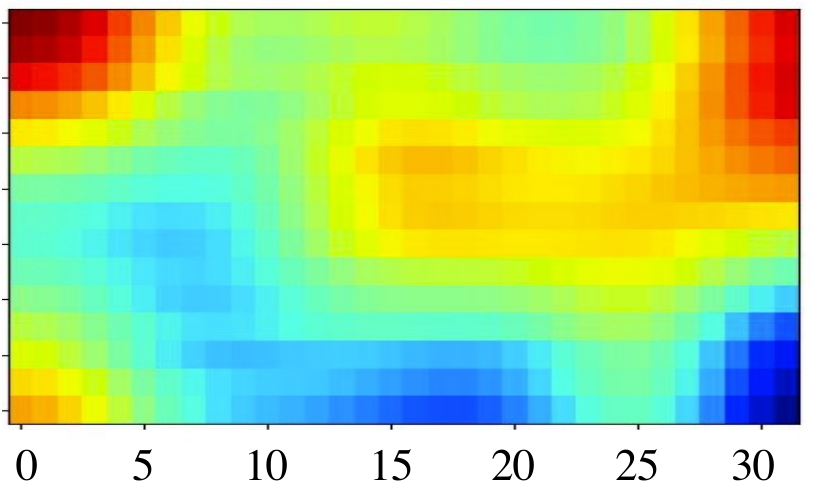}
}
\subfigure[Correlated (radial, $L=$~4)]{
\includegraphics[width=.25\textwidth]{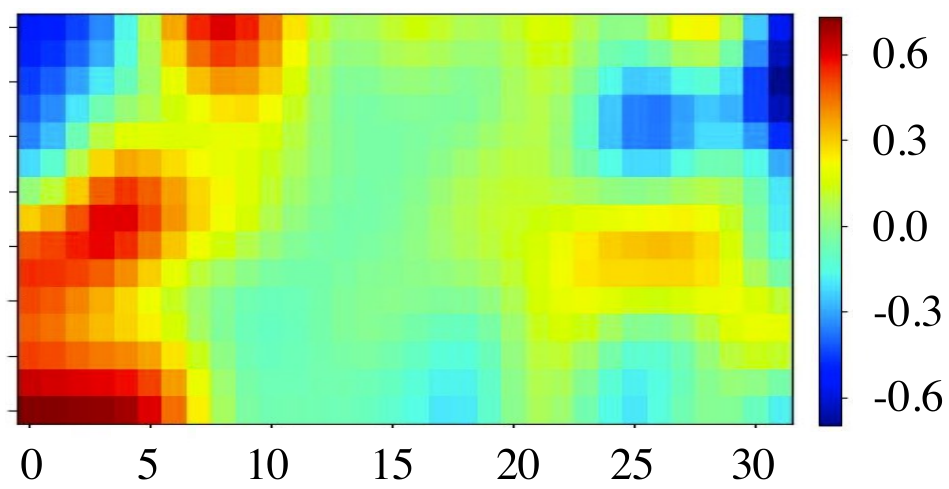}
}
\caption{Generating spatially-correlated variation maps: Uncorrelated variation maps ((a)-(b)), generated from a Gaussian distribution, are subsequently convolved with a Gaussian kernel (c) to obtain correlated variation maps ((d)-(g)).}
\label{fig7}
\end{figure*}

 Recall that the weight matrices in the IPNN case study are 16$\times$16 (input layer), 16$\times$16 (first hidden layer), and 10$\times$16 (second hidden layer). Therefore, post SVD, the MZI arrays are used to represent five 16$\times$16 and one 10$\times$10 unitary matrices. To apply correlated uncertainties to these MZI arrays, we consider the MZIs to be arranged in a grid (see Fig. \ref{fig6}). For any $N\times N$ unitary matrix, it can be shown that the MZIs can be arranged in a $(N-1)$ units $\times$ $2N$ units grid, where one unit corresponds to half the length of an MZI. Fig. \ref{fig6} shows such a grid for a 16$\times$16 unitary matrix. Therefore, to analyze the impact of correlated uncertainties, we create $(N-1)$ units $\times$ $2N$ units variation maps. Fig. \ref{fig7}(a) shows an example of an uncorrelated 15$\times$32 variation map with $\sigma_{PhS}=$~0.025. Note that this is an approximate model where we assume that MZIs that are topologically adjacent will also be physically adjacent in the floorplan to minimize the waveguide length. However, our approach will remain applicable if the layout of MZIs differs from the array topology to satisfy routing and design-rule check constraints. Details on how different variation maps are generated can be found in the Appendix.\par

\begin{figure*}[t]
\subfigure[]{
\includegraphics[width=.31\textwidth]{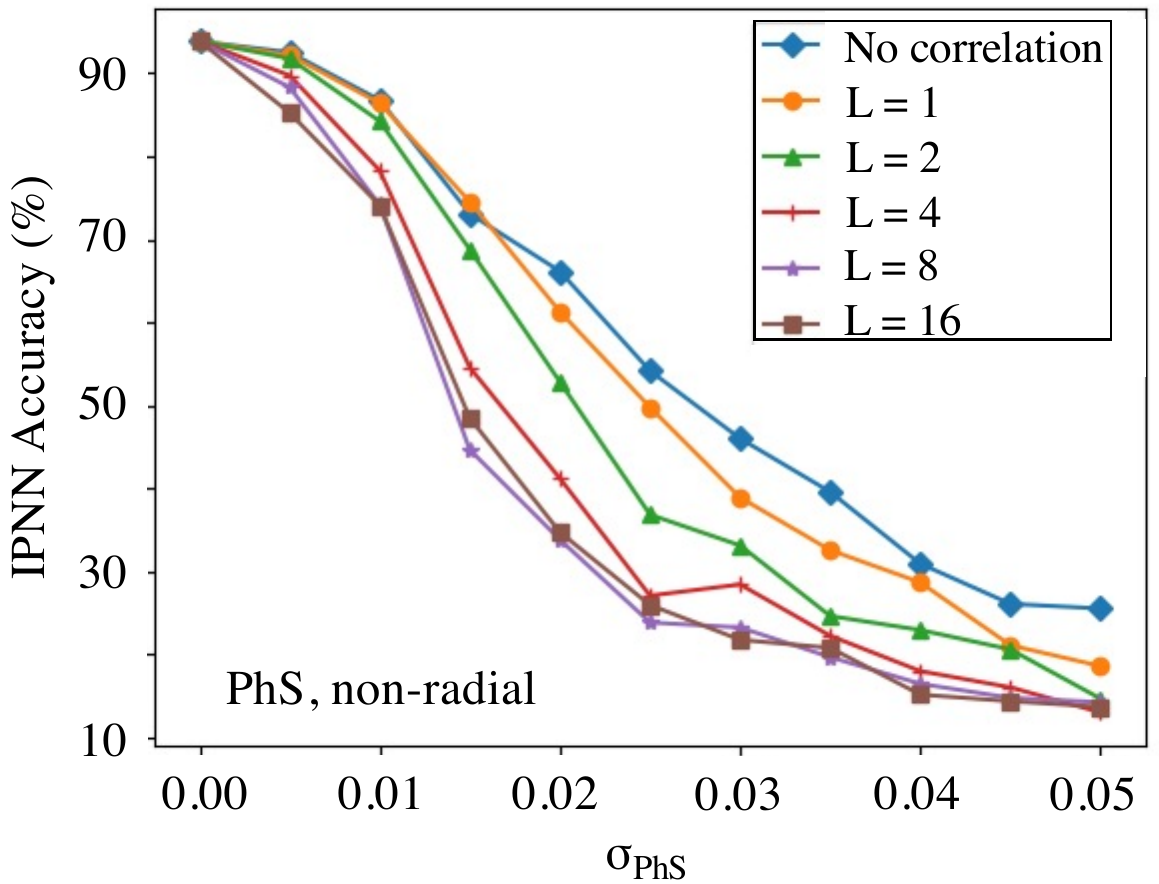}
}
\subfigure[]{
\includegraphics[width=.31\textwidth]{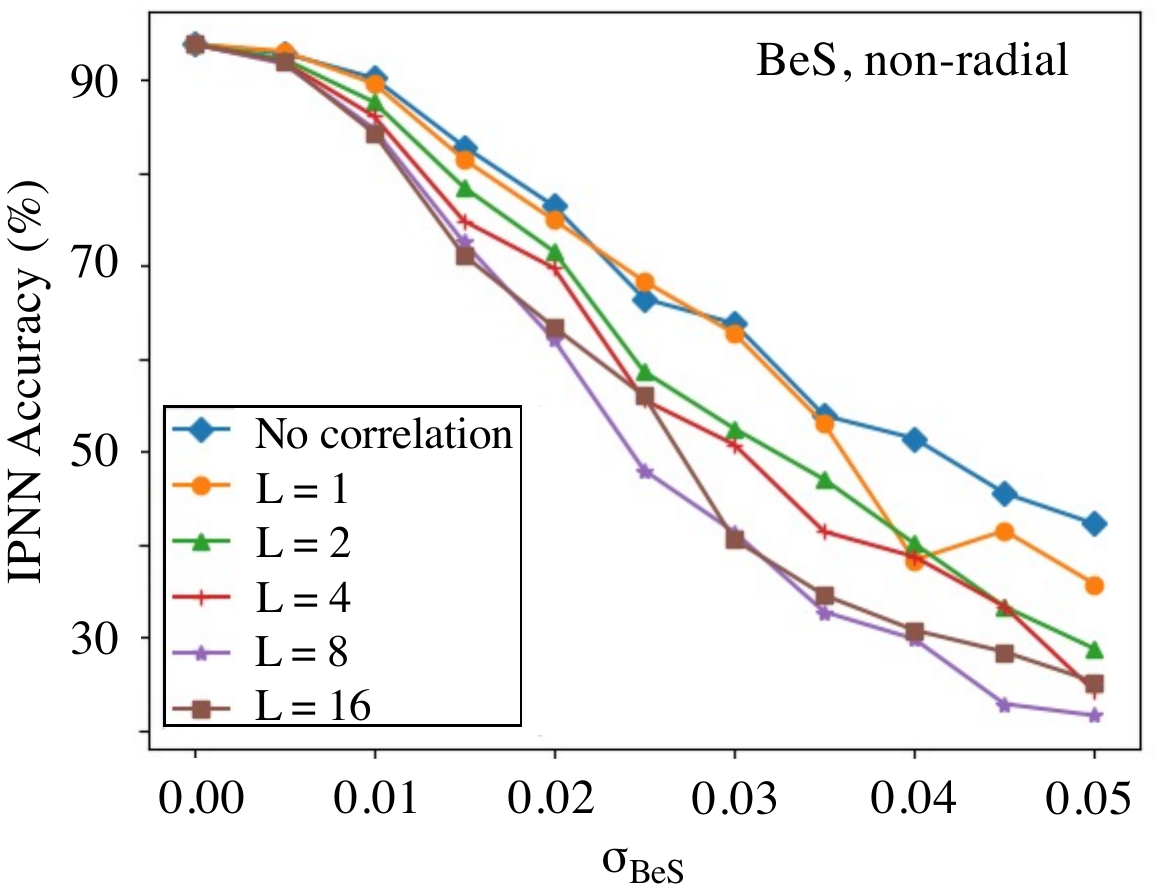}
}
\subfigure[]{
\includegraphics[width=.31\textwidth]{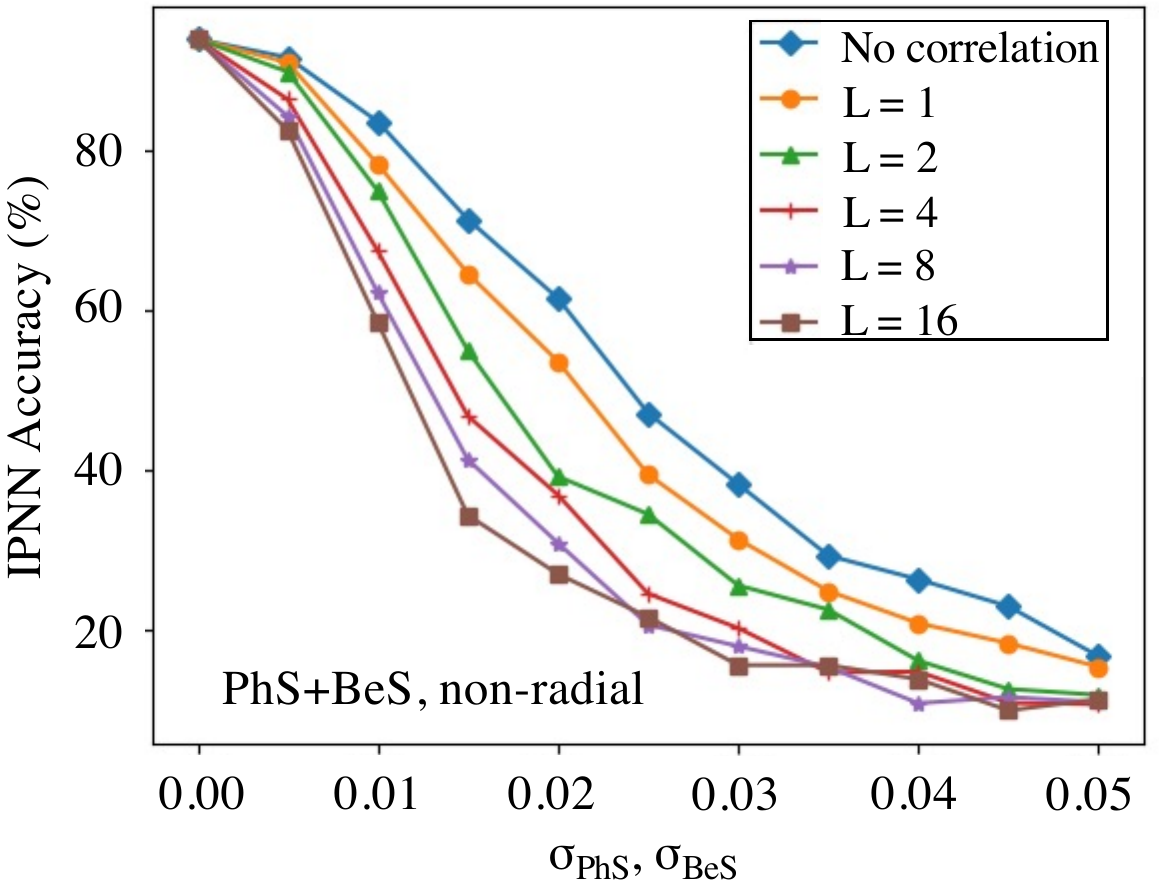}
}
\newline
\subfigure[]{
\includegraphics[width=.31\textwidth]{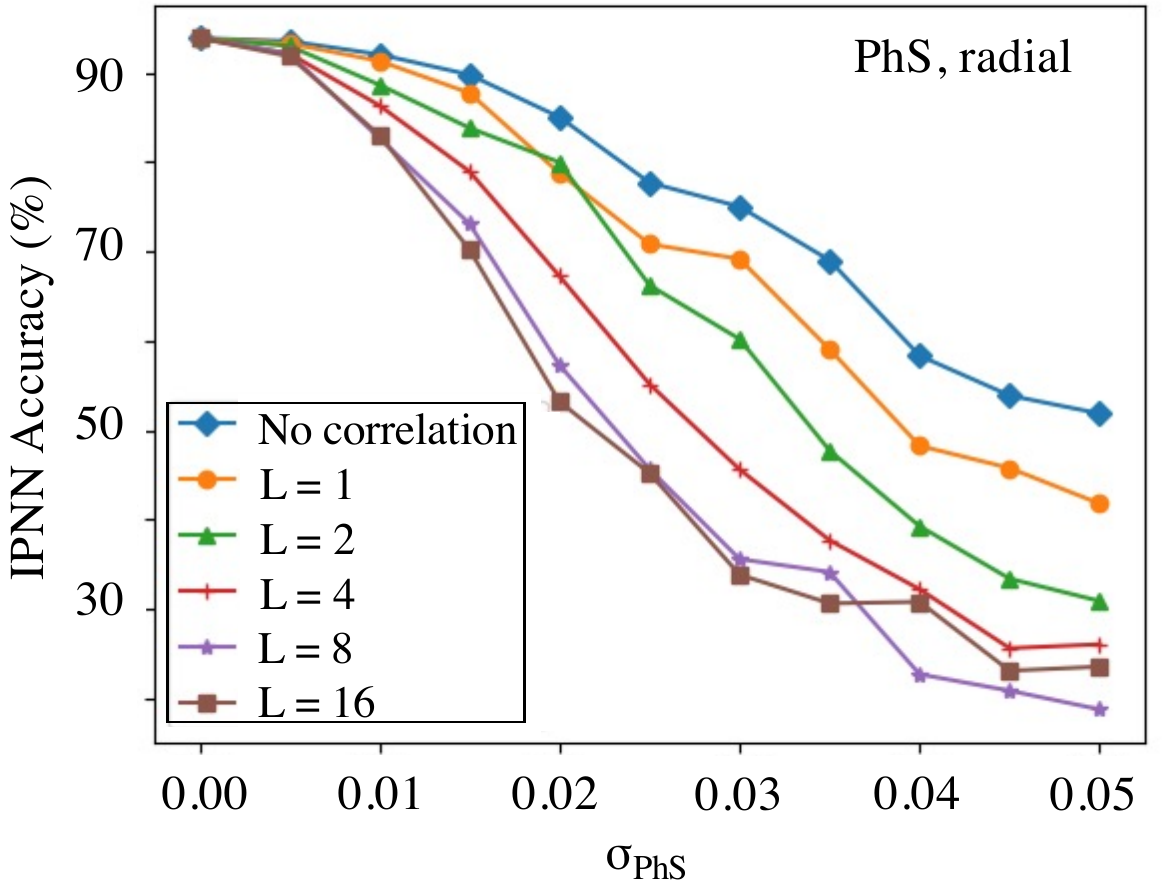}
}
\subfigure[]{
\includegraphics[width=.31\textwidth]{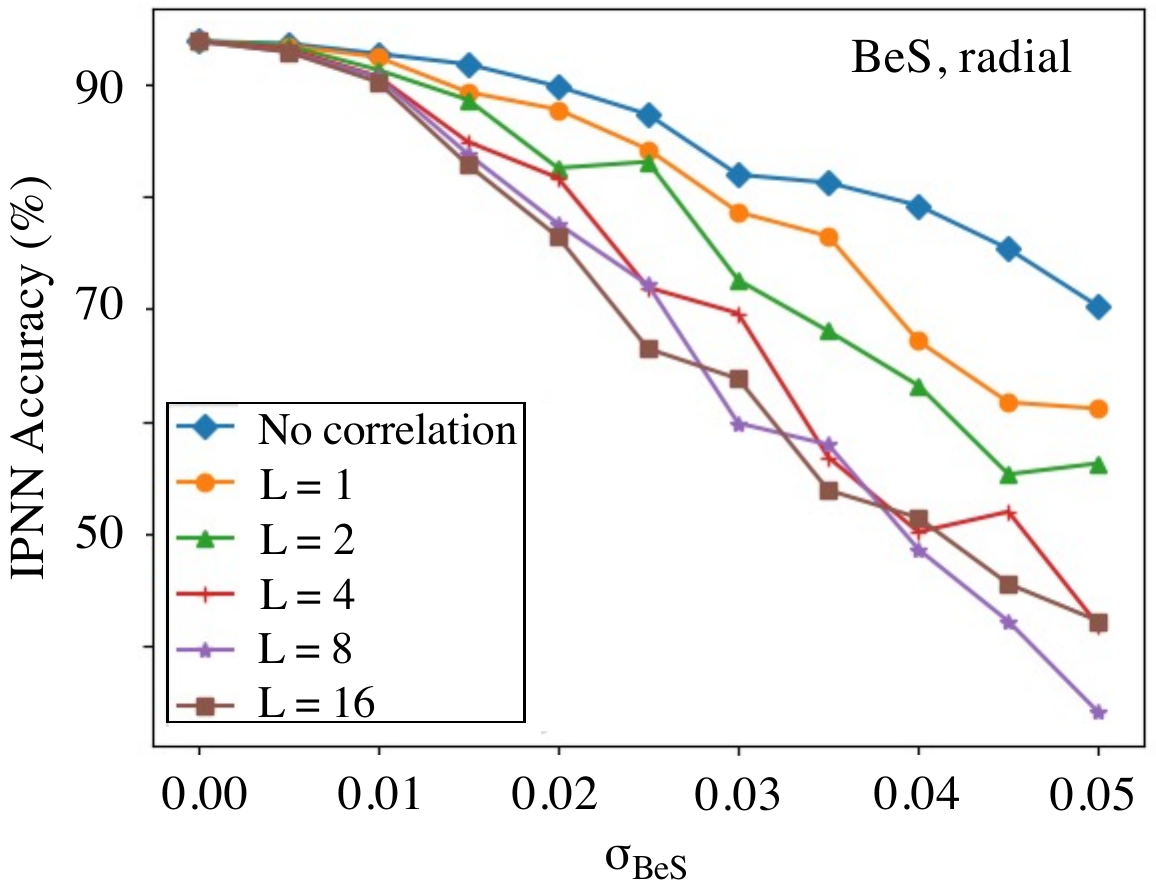}
}
\subfigure[]{
\includegraphics[width=.31\textwidth]{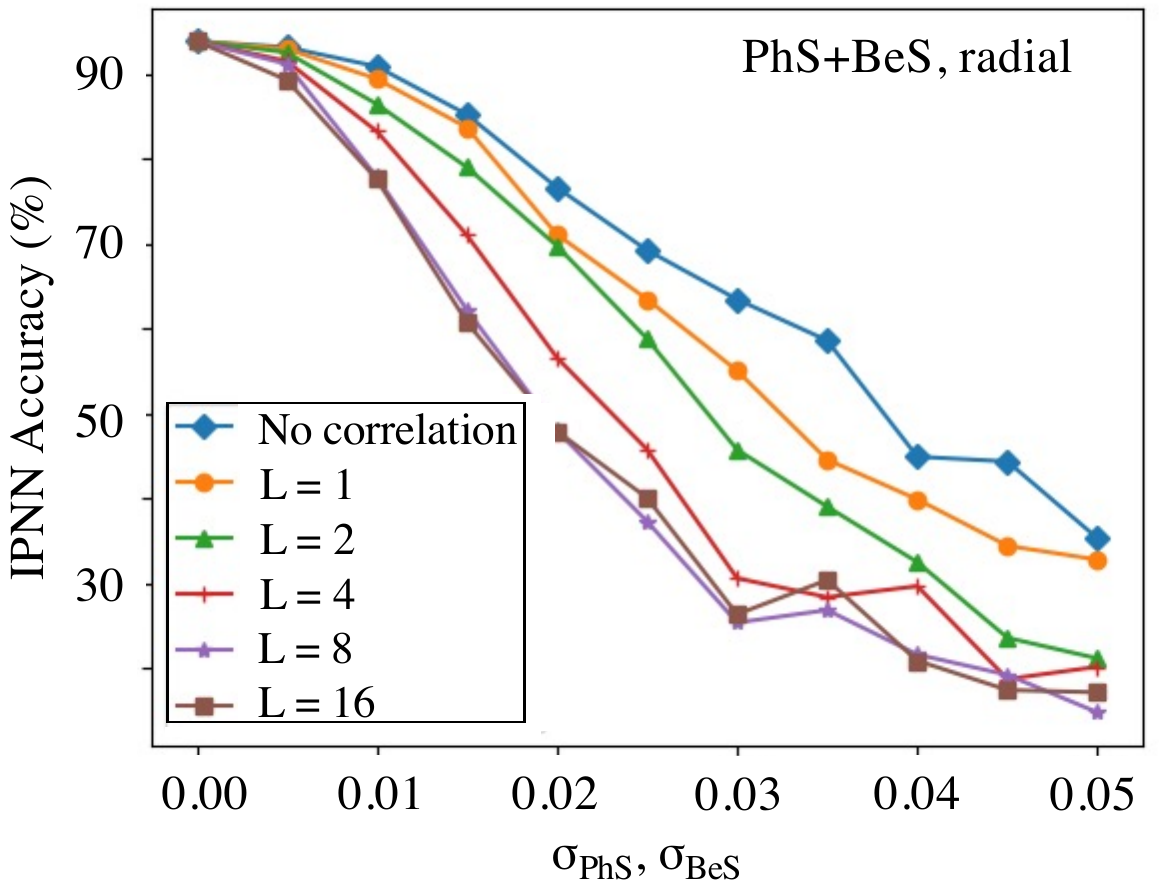}
}
\caption{Impact of spatially correlated non-radial and radial uncertainties on IPNN inferencing accuracy. Here, $L$ denotes the correlation length defined in units. One unit corresponds to $l_{MZI}/2$, where $l_{MZI}$ denotes the length of an MZI.}
\label{fig8}
\vspace{-1.5em}
\end{figure*} 


Prior experimental efforts (e.g, \cite{mirza2021silicon} and \cite{su2014reduced}) have reported that for some uncertainties---especially those due to FPVs---the non-uniformity increases radially from the center of a wafer towards its edges (i.e., the wafer center experiences less variations). Fig. \ref{fig7}(b) shows one such randomly generated 15$\times$32 uncorrelated radial variation map with $\sigma_{PhS}=$~0.025. Figs. \ref{fig7}(d)--(f) show the spatially correlated variation maps obtained when the uncorrelated variation map (Fig. \ref{fig7}(a)) is convoluted with Gaussian kernels with $L=$~2, 4, and 8 units, respectively. Observe that as $L$ increases, the uncertainties in the variation maps are spread out over larger contiguous areas. Similarly, Fig. \ref{fig7}(g) shows the correlated variation map when the uncorrelated radial variation map (Fig. \ref{fig7} (b)) is convoluted with a Gaussian kernel with $L=4$. Observe also that, as expected, regions towards the periphery of the correlated radial variation maps are more prone to experience higher levels of uncertainties (red and blue regions in Fig. \ref{fig7}(g)). \par
 
The uncertainty associated with each 1 unit~$\times$~1 unit block in a correlated variation map is applied to the phase angle (or splitting ratio or both) of the corresponding MZI and the IPNN inferencing accuracy is obtained. This procedure is repeated for 1000 iterations to obtain the mean inferencing accuracy under correlated variations. Figs. \ref{fig8}(a)-(c) show the mean IPNN inferencing accuracy when spatially correlated non-radial uncertainties with different $L$s are introduced in PhS, BeS, and both, respectively. We observe that in all cases, under similar $\sigma$, the accuracy is lower when the uncertainties are spatially correlated, compared to the uncorrelated scenario. Note also that as $L$ increases, the accuracy degrades further in all cases. This is an interesting observation given prior studies on correlated uncertainties in integrated photonic circuits for Datacom applications (e.g., \cite{mirza2021silicon}, \cite{lu2017performance}) that showed the impact of uncertainties on a system performance becomes less catastrophic as the correlation length increases. In other words, a higher correlation length results in better frequency matching (i.e., inter-device matching) among apart devices (e.g., microring resonators in \cite{mirza2020variation}). Nevertheless, coherent IPNNs based on MZIs do not necessarily benefit from inter-device matching. Highly correlated uncertainties in IPNNs lead to a large deviation in the IPNN parameters in one direction. For example, all PhS in an extended region (based on the large correlation length) may experience a positive shift in the phase angles. As a result, the optical signal traversing a cascade of deviated PhS in MZIs experiences a larger undesired phase shift, resulting in higher performance degradation. Conversely, in the presence of more localized uncertainties (i.e., shorter correlation lengths), parameters in cascaded MZIs will likely experience deviations in different directions and will likely cancel each other (e.g., a positive phase shift on an optical signal cancels a prior negative phase shift). Therefore, such localized uncertainties will potentially have a lower impact on IPNN accuracy, as shown by the blue lines in Figs. \ref{fig8}. 



This trend demonstrates an interesting aspect of random uncertainties in IPNNs. The uncorrelated variation maps shown in Figs. \ref{fig7}(a)-(b) have a few MZIs with extremely high levels of uncertainties (hotspots), while most other MZIs have very low variations. Upon convolution with the Gaussian kernel, which is a smoothing filter, the high levels of variations at these hotspots are distributed across adjoining blocks. Consequently, while we have very few (if any) hotspots in the correlated variation maps, many more blocks have somewhat high levels of variations. Our simulation results show that given two distributions of uncertainties, one where a few MZIs experience extreme uncertainties (low correlation length) and another where several MZIs experience moderate-to-high uncertainties (high correlation length) ---the latter is expected to have more impact on the IPNN accuracy. Therefore, efforts to improve the tolerance of IPNNs should consider addressing the uncertainties in all the MZIs in an array, rather than being targeted towards a select few MZIs with catastrophic uncertainties. \par

Figs. \ref{fig8}(d)-(f) show that the aforementioned trend also holds for spatially correlated radial uncertainties. Observe also that for similar $\sigma$, the IPNN accuracy is higher for radial uncertainties compared to non-radial uncertainties. Recall that in the radial case, the magnitude of uncertainties experienced by each block in the array is scaled based on its Euclidean distance from the center of the grid. Therefore, only the MZIs at the grid edges have $\sigma \approx \sigma_{PhS}$, while for all other blocks $\sigma < \sigma_{PhS}$. In contrast, for non-radial cases, all MZIs in the grid have $\sigma = \sigma_{PhS}$. As a result, every MZI in a radial variation map is expected to have smaller (or at most equal) uncertainties compared to the corresponding MZI in a non-radial variation map.

%% file: systematic.tex
In addition to uncertainties in phase angles and splitting ratios, MZIs are prone to imprecisions due to inevitable optical losses and quantization errors introduced in the phase angles when they are encoded using low precision DACs. In this section, we present an analytical model to capture the impact of optical losses in IPNNs and, for the first time, quantify IPNN performance degradation when the optical losses in different MZIs have different statistical distributions (different standard deviations). We also identify different ways in which phase angles can be encoded and simulate the impact of quantization errors on the IPNN accuracy. 

\subsection{IPNNs under Non-Uniform MZI Insertion Loss}
\label{loss}
A drawback of coherent IPNNs is the increased chip size due to the bulky MZIs \cite{shibata2008compact}. The large footprint of MZIs can be attributed to the relatively small thermo-optic coefficient of most opto-electronic materials. As a result, an optical path length of hundreds of micrometers is necessary for a required phase change \cite{deng2021parity}. In fact, state-of-the-art MZIs, such as the one proposed in \cite{shokraneh2020diamond}, are as long as $\approx$300$\mu$m, with the majority of the footprint occupied by the two phase shifters (each up to 135 $\mu$m long). Optical signal traversing such bulky MZIs can experience optical losses in the 3-dB BeS, absorption loss due to the microheaters' metal planes, and propagation loss in the waveguides. Prior work showed that a total insertion loss of up to 1.2~dB per MZI is expected, even for state-of-the-art MZIs \cite{shibata2008compact}. During the training of IPNNs (in software), we consider lossless MZIs. However, the on-chip insertion loss can lead to a degraded inferencing accuracy. 

Consider an MZI with two BeS (see Fig. \ref{placeholder}). The transfer matrix of a lossless 3-dB beam splitter is given by (\ref{E01}) with $r_{00}=r_{11}=t_{01}=t_{10}=\frac{1}{\sqrt{2}}$. Suppose in the input splitter the optical signal amplitude in the top arm is attenuated by a factor of $\beta_{lt}$ while the one in the bottom arm is attenuated by $\beta_{lb}$. The transfer matrix of this lossy beam splitter is:
\begin{equation}
   \begin{pmatrix} \beta_{lt}/\sqrt{2} & i\beta_{lb}/\sqrt{2}  \\ i\beta_{lt}/\sqrt{2} & \beta_{lb}/\sqrt{2}  \end{pmatrix}. 
   \label{lossyBeS}
\end{equation}
Similarly, note that the attenuation factors in the top and bottom arms of the output splitter are given by $\beta_{rt}$ and $\beta_{rb}$, respectively. The transfer matrix of the lossy MZI is given by:
\begin{equation}
    \tilde{T}_{MZI}(\theta, \phi)=\begin{pmatrix}
    \frac{\beta_{rt}\beta_{lt}e^{i(\theta+\phi)}-\beta_{rb}\beta_{lt}e^{i\phi}}{2} & \frac{i\beta_{rt}\beta_{lb}e^{i\theta}+i\beta_{rb}\beta_{lb}}{2} \\
    \frac{i\beta_{rt}\beta_{lt}e^{i(\theta+\phi)}+it\beta_{rb}\beta_{lt}e^{i\phi}}{2} & \frac{-\beta_{rt}\beta_{lb}e^{i\theta}+\beta_{rb}\beta_{lt}}{2}
    \end{pmatrix}.
    \label{T_MZI_lossy}
\end{equation}
In a special case when $\beta_{lt}=\beta_{lb}=\beta_{rt}=\beta_{rb}=\beta$, $\tilde{T}_{MZI}=\beta^{2}T_{MZI}$. This overall amplitude attenuation of $\beta^{2}$ corresponds to a power attenuation of $\beta^{4}$. Therefore, the insertion loss, in dB, is given by $IL=$~10~$\log_{10} \beta^4$. Note that the loss modeled for an MZI (in terms of $\beta$) represents the total MZI insertion loss, including the waveguide propagation loss, PhS metal absorption loss, and BeS coupling loss). 

\begin{figure}[t]
  \centering
  \includegraphics[scale=0.49]{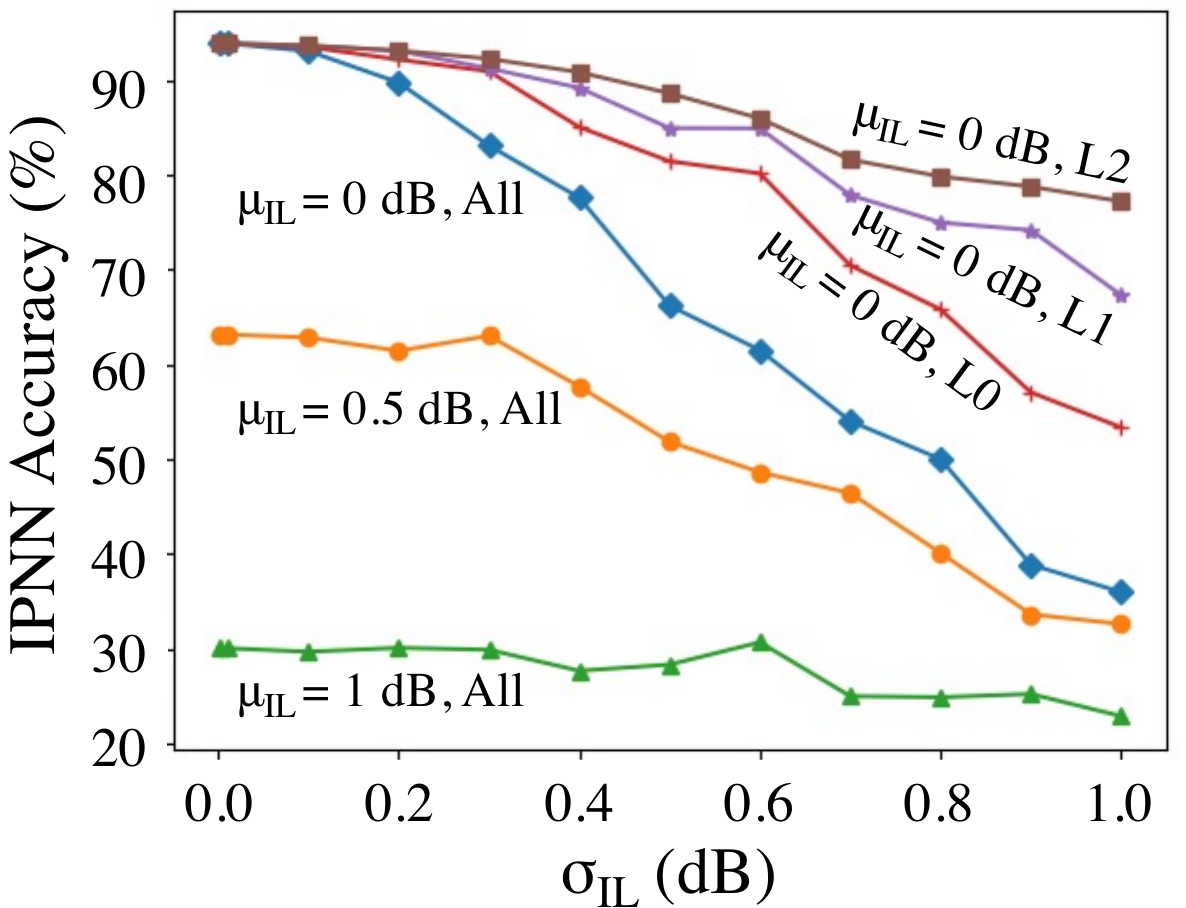}
  \caption{Impact of MZI insertion loss on the IPNN accuracy.}
  \vspace{-1.6em}
  \label{fig9}
\end{figure}

To analyze the impact of non-uniform MZI insertion loss on the IPNN inferencing accuracy, we perform 1000 Monte Carlo iterations. In each iteration, the insertion loss for each MZI is sampled from a Gaussian distribution, $IL=\mu_{IL}+\mathcal{N}\left(0,\sigma_{IL}^2\right)$. We take the mean of the 1000 classification accuracies as the final classification accuracy. Fig. \ref{fig9} shows the mean inferencing accuracy in the IPNN case study in the presence of lossy MZIs. We observe that when MZIs in all layers are lossy, the accuracy degrades significantly with increasing $\mu_{IL}$. Even when $\mu_{IL}=$~0 dB, the accuracy falls sharply with increasing $\sigma_{IL}$. This shows that increasing non-uniformity in the insertion loss across MZIs (i.e., with increasing $\sigma_{IL}$) can be catastrophic to the IPNN performance, even when the mean insertion loss is zero. To highlight the impact of insertion loss in different layers, we consider lossy MZIs in the three hidden layers (L0, L1, and L2), one layer at a time (see Fig. \ref{fig9}). When insertion loss is introduced in one layer, the MZIs in the other layers are assumed to be lossless. We observe that for similar levels of insertion loss, the IPNN accuracy is the least when MZIs in L0 (first hidden layer) are lossy, followed by when MZIs in L1 and those in L2 are lossy. This trend can be attributed to the fact that errors due to insertion loss in the first hidden layer affect the matrix multiplication in all subsequent layers and are propagated across the highly interconnected MZI array, thereby highly degrading the IPNN accuracy. Alternatively, errors introduced due to insertion loss in the final layer are not propagated widely and have a relatively localized impact. \par

\begin{figure*}[t]
\centering

\subfigure[Equidistant voltage steps (EVS)]{
\includegraphics[width=.26\textwidth]{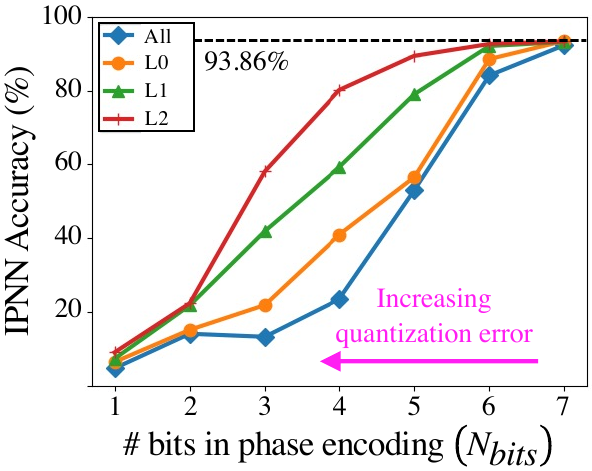}
}
\subfigure[Equidistant phase steps (EPS)]{
\includegraphics[width=.22\textwidth]{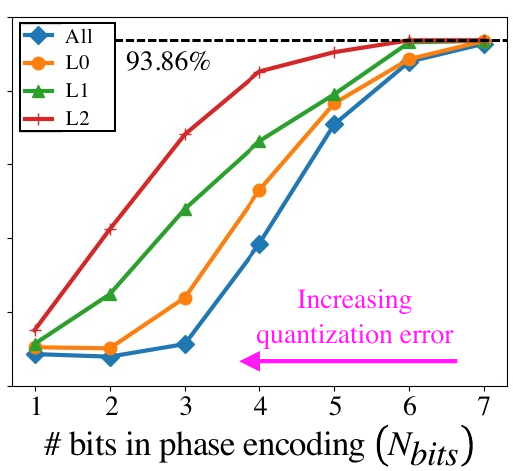}
}
\subfigure[K-means clustering (KC)]{
\includegraphics[width=.22\textwidth]{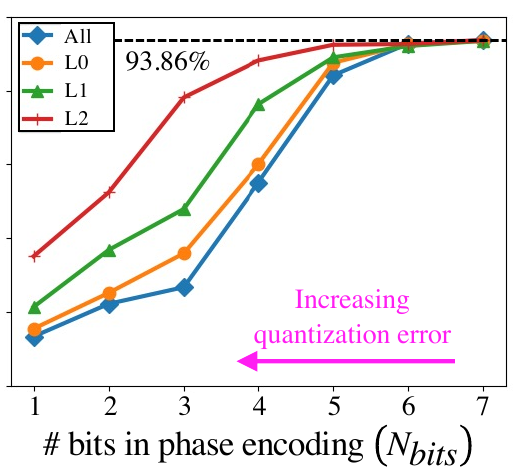}
}
\subfigure[Comparison]{
\includegraphics[width=.22\textwidth]{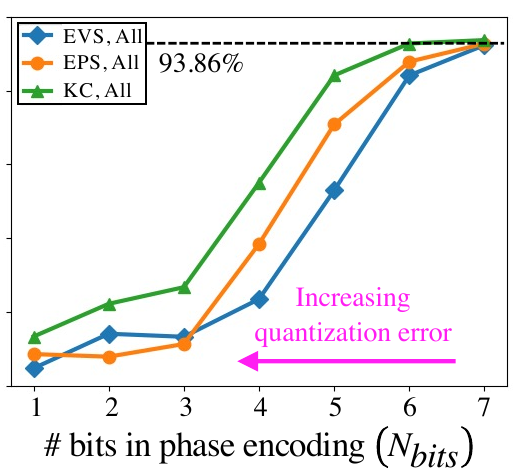}
}
\caption{The impact of different phase encoding approaches on the IPNN inferencing accuracy under low-precision settings.}
\label{fig10}
\vspace{-1em}
\end{figure*}

From the above discussion, it is clear that the IPNN classification accuracy is sensitive to MZI insertion loss. Therefore, reducing the MZI loss is critical, especially for ultra-deep IPNNs trained for more complex tasks. Moreover, even for a low $\mu_{IL}$, the IPNN accuracy decreases with increasing non-uniformity ($\sigma_{IL}$) in the insertion loss of different MZIs. This indicates that FPVs, which lead to such non-uniformity in the loss, also need to be minimized. We have also shown that insertion loss in the initial (i.e., closer to the input) IPNN layers has a high impact on the IPNN inferencing accuracy. To compensate for the MZI insertion loss, we should ideally include additional semiconductor optical amplifiers (SOAs) after each layer in the IPNN. However, due to strict area constraints, we may be able to afford amplification only after selected layers. Our analysis shows that, in such cases, SOAs should preferentially be placed after the initial IPNN layers.

%% file: precision.tex
Thermo-optic PhS utilize the high thermo-optic coefficient of silicon to apply tunable phase shifts to the optical signal traversing a waveguide. Recall from Section \ref{random}.A that the temperature-dependent phase change in PhS is given by \eqref{del_phi}. The microheaters in PhS are controlled by applying a tuned voltage across the heater coil. The voltage is supplied from a direct current (DC) source based on a digital-to-analog converter (DAC). The precision of the temperature shift $\Delta T$, and in turn, the phase angle is limited by the quantization error in the DAC. The magnitude of this quantization error increases with decreasing bit-precision in a DAC. The power consumption of a DAC with a precision of $N_{bits}$ is proportional to $\frac{2^{N_{bits}}}{(N_{bits}+1)}$ \cite{saberi2011analysis}. Therefore, low-precision DACs, which are susceptible to quantization errors, are often required for low-power applications. In addition, phase angles encoded with lower $N_{bits}$ can also help lower memory overhead. \par


Recall from (\ref{precision_del_phi}), for a phase shift of $\Phi$ we have $\Phi=K\cdot V^2$, where $K$ is a constant of proportionality that depends on the structure of the phase shifter. The voltage required for a phase shift of $\pi$, given by $V_{\pi}=\sqrt{\frac{\pi}{K}}$, is used as a figure of merit for PhS. Among existing PhS, the one proposed in \cite{harris2014efficient} offers the minimum of $V_{\pi}=$~4.36~V, and has been considered for the simulations in this paper. With $V_{\pi}=$~4.36 V, we use \eqref{precision_del_phi} to obtain $K=$~0.165. Therefore, to realize phase angles 0~$\leq \phi \leq$~2$\pi$ in an MZI, we need a voltage 0~$\leq V \leq$~6.166 V. For an encoding with $N_{bits}$ bits in the voltage driver, we can have $2^{N_{bits}}$ voltage intervals in the range [0, 6.166 V]. The voltage drivers (see Fig. \ref{placeholder}(d)) can assign the input voltage ($V$) at these $2^{N_{bits}}$ steps in one of following three ways:
\vspace{-0.2em}
\begin{itemize}
    \item Equidistant voltage steps (EVS): $2^{N_{bits}}$ equidistant voltage steps are considered in the range [0, 6.166 V] with a step-size of $\frac{6.166}{(2^{N_{bits}}-1)}$ V. In this case, the voltage driver simply amplifies the output from the DAC. While EVS is easier to implement, we will show that it underperforms for low $N_{bits}$.    
    \item Equidistant phase steps (EPS): $2^{N_{bits}}$ equidistant phase steps are considered in the range [0, 2$\pi$] with a step size of $\frac{2\pi}{(2^{N_{bits}}-1)}$ radians. The voltages corresponding to these $2^{N_{bits}}$ phase values are considered as the $2^{N_{bits}}$ voltage steps in the driver and are \textit{not} equidistant. In this case, the voltage driver needs to perform additional computation on the DAC output to generate the voltage supplied to the microheater.
    \item K-means clustering (KC): Using K-means clustering, we divide the weights in the IPNN into $2^{N_{bits}}$ clusters. The voltage values corresponding to the median phase angle of each cluster constitute the $2^{N_{bits}}$ voltage steps. Therefore, the voltage driver needs to supply the voltage corresponding to the median phase angles for each of the $2^{N_{bits}}$ clusters and therefore, requires additional computation and memory (to store the median phase angles).   
\end{itemize}

Note that the above finite-precision settings are applied during inferencing; we assume that the software training has been performed in full precision. Fig. \ref{fig10} shows the IPNN inferencing accuracy under these finite-precision settings. We observe that for both EVS and EPS, IPNN achieves the full-precision accuracy (93.86\%) with $N_{bits}=$~7. This signifies that we require only 128 voltage steps in the range [0, 6.166]~V to maximize IPNN accuracy. Alternatively, with KC, we can achieve full-precision accuracy with just $N_{bits}=$~6. Fig. \ref{fig10}(d) compares the three settings and shows that KC offers the maximum accuracy at low precision. This is because it takes the distribution of the trained weights into account. However, the voltage steps obtained from KC are unique to each trained IPNN. Therefore, KC involves additional computation and hardware overhead. Note also that for a given $N_{bits}$, EPS typically offers a higher accuracy compared to EVS. We know from (\ref{precision_del_phi}) that the phase angles are proportional to $V^2$. Therefore, in EVS, the gap between the phase steps (corresponding to the equidistant voltage steps) increases with the phase angle. Consequently, the higher phase angles (more important weights) have a higher quantization error. In contrast, all phase angles have similar quantization errors in EPS, thereby leading to a higher accuracy under low precision. \par

We also analyze the performance of EVS, EPS, and KC when they are applied to an individual layer in the IPNN. Recall that we obtain full-precision accuracy (93.86\%) for $N_{bits}\geq 7$. Therefore, when we vary $N_{bits}$ for one layer, all the weights in the remaining layers are quantized using $N_{bits}=$~8; this ensures that the loss in accuracy is solely due to quantization errors in the layer of interest. We observe that for all the three methods, the reduction in accuracy is maximum when low-precision settings are applied to the initial (i.e., closest to the input) layer L0. Recall from Section \ref{loss} that we have a similar observation for MZI insertion loss: the IPNN accuracy is affected the most when MZIs in the initial layers are lossy. Both these observations are likely due to the same reason as errors due to lossy MZIs and low-precision PhS in initial layers can spread across the MZI arrays, thereby leading to a higher accuracy loss. \par

In summary, even though IPNNs can offer high accuracy with just a 7-bit quantized DAC, they are sensitive to low-precision settings. KC can be used to improve the accuracy, but it involves additional computation and hardware overhead. Moreover, PhS in the initial layers are more prone to the catastrophic impact of low-precision phase encoding.  

%% file: case_study.tex
In practice, IPNNs will likely encounter multiple uncertainties simultaneously while experiencing optical loss and quantization errors. Sections \ref{random} and \ref{imprecision} comprehensively explore the \textit{impact} of each of the different imperfections considered in this paper, including uncertainties in phase angles and splitting ratios, non-uniform MZI insertion loss, and low-precision phase encoding. In this section, we present a case study of a representative IPNN and show how its inferencing accuracy deviates under different situations of multiple simultaneous imperfections. We show that the loss in inferencing accuracy in each situation cannot be predicted by simply superposing the impact of each standalone imperfection (i.e., the accuracy losses under standalone imperfections are not additive). As a result, an understanding of IPNN reliability necessitates a detailed case study similar to the one we present in this section. \par

For this case study, we apply imperfections in the three-layer (one input and two hidden) feed-forward IPNN described in Section \ref{random}.D and measure its inferencing accuracy for different imperfection parameter sets. Next, we define some terms that will help us represent the simultaneous imperfections and quantify their impact on the IPNN inferencing accuracy.
\begin{definition}
    An \textbf{imperfection parameter set} $P=\{\sigma_{PhS}, \sigma_{BeS}, L, \sigma_{IL}, N_{bits}\}$ is defined as the ordered quintuplet of the parameters quantifying the different IPNN imperfections under which inferencing is performed. 
\end{definition}
For a given $P$, the correlation length $L$ and the phase-encoding $N_{bits}$ are deterministic, while $\sigma_{PhS}$, $\sigma_{BeS}$, and $\sigma_{IL}$ describe the distribution of uncertainties in the phase angles, uncertainties in the splitting ratios, and non-uniformity in the MZI insertion loss, respectively. Note that, in the interest of brevity, we only consider radial uncertainties phase angles and splitting ratios. To obtain the accuracy loss for a particular $P$, we perform Monte-Carlo simulations by randomly sampling several imperfection instances.  
\begin{definition}
    For a given $P=\{\sigma_{PhS}, \sigma_{BeS}, L, \sigma_{IL}, N_{bits}\}$ and a trained IPNN, an \textbf{imperfection instance} $p$ is defined as a version of the IPNN where in each MZI
    \begin{itemize}
        \item the phase angles and splitting ratios are perturbed based on variation maps described by $u_{PhS}(x,y, L, \sigma_{PhS})$ and $u_{BeS}(x,y, L, \sigma_{BeS})$, respectively;
        \item the insertion loss is sampled from a Gaussian distribution $\mathcal{N}\left(0,\sigma_{IL}^2\right)$; and
        \item the phase angles are encoded as $2^{N_{bits}}$ equidistant voltage steps in the range [0, $V_{\pi}$].
    \end{itemize}
\end{definition}
\noindent Here, $(x,y)$ denotes the position of the PhS or BeS (see Section \ref{random}.D.3 and Fig. \ref{fig6}), and $u_{PhS}$ and $u_{BeS}$ denote the correlated variation maps for phase angles and splitting ratios, respectively (see Appendix). Also, recall from Section \ref{imprecision}.B that $V_{\pi}$ for a phase shifter denotes the voltage from a direct current (DC) source required for a phase shift of $\pi$. Across all our simulations, for a given $P$, we perform inferencing using $n_p=$~10 randomly generated $p$'s and obtain the mean inferencing accuracy. The \textit{\textbf{simulated accuracy loss}}, $SAL_{10}(P)$, is then given by the difference between the nominal IPNN accuracy, which is 93.86\% in our case study, and the mean inferencing accuracy under imperfections. We also define the aggregated accuracy loss for a $P$ as follows.



\begin{figure}[t]
 \centering
   \subfigure[]
   {\includegraphics[width=0.48\textwidth]{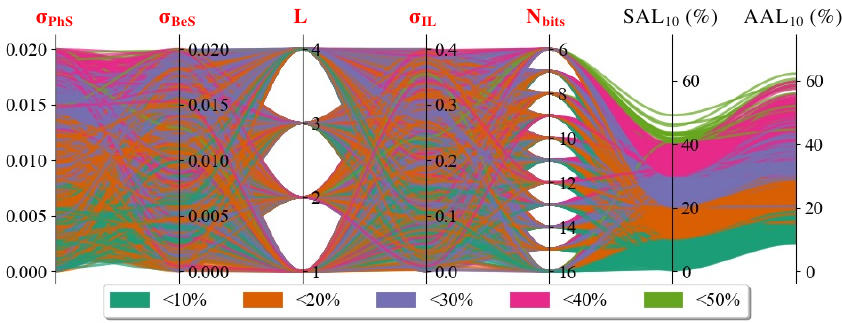}\label{baseline}}
  \subfigure[]{
  \includegraphics[width=0.48\textwidth]{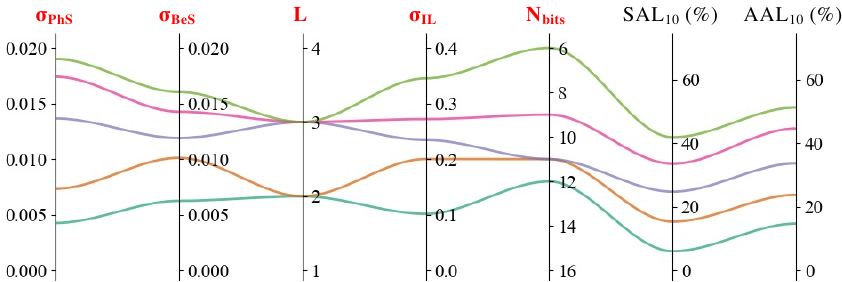} \label{ltpplots}
  } \vspace{-0.05in}
  \caption{(a) $SAL_{10}$ and $AAL_{10}$ for 2000 imperfection parameter sets ($P$s) where each imperfection parameter varies over a wide range. The curve corresponding to each parameter set is color-coded based on its $SAL_{10}$ (see legend). (b) Median of the imperfection parameters, $SAL_{10}$, and $AAL_{10}$ for the imperfection parameter sets in each $SAL_{10}$ band in (a).}
 \vspace{-0.3in}
 \label{allvarying}
 \end{figure}

\begin{definition}
    For a given $P$, the \textbf{aggregated accuracy loss} for an IPNN is given by: 
\end{definition}
\vspace{-2em}
\begin{equation}\label{AAL_np}
    \begin{aligned}
    & AAL_{n_p}(P)=SAL_{n_p}(\{\sigma_{PhS}, 0, 0, 0, 0\}) \\
    & \quad +SAL_{n_p}(\{0, \sigma_{BeS}, 0, 0, 0\})+SAL_{n_p}(\{0, 0, L, 0, 0\}) \\
    & \quad +SAL_{n_p}(\{0, 0, 0, \sigma_{IL}, 0\})+SAL_{n_p}(\{0, 0, 0, 0, N_{bits}\}).
    \end{aligned}
\end{equation}

In (\ref{AAL_np}), $AAL_{n_p}(P)$ is simply the sum of the standalone $SAL_{n_p}$ of each imperfection parameter. In Fig. \ref{allvarying}(a), we use a parallel coordinates plot with cubic Bezier curves to compare the $SAL_{10}$ and $AAL_{10}$ for 2000 different imperfection parameter sets. Each curve in the plot corresponds to an imperfection parameter set and is color-coded based on its $SAL_{10}$ (see legend in Fig. \ref{allvarying}). The vertical axes in the plot are oriented such that a higher intercept on an axis signifies a larger deviation from the ideal IPNN parameters (hence, leading to a higher $SAL_{10}$). Therefore, the values of $\sigma_{PhS}$, $\sigma_{BeS}$, $L$, $\sigma_{IL}$, $SAL_{10}$, and $AAL_{10}$ increase from bottom to top, while the reverse is true for $N_{bits}$ (recall that a lower number of bits in phase encoding leads to a higher $SAL_{10}$). Note also that we limit the range for each imperfection parameter to a subset of their respective ranges in the Sections \ref{random}-IV. This ensures that we have enough imperfection parameter sets ($P$'s) where the $SAL_{10}$ is not so high that the inferencing accuracy of the IPNN is less than the probability of a random choice (10\% for the MNIST dataset). However, our analysis can be easily extended to any user-defined range of the imperfection parameters. From Fig. \ref{allvarying}(a), we observe up to a 46\% simulated accuracy loss ($SAL_{10}$) due to simultaneous imperfections within the parameter ranges considered. We also find that for all $P$'s, $AAL_{10}(P)\geq SAL_{10}(P)$, i.e., the sum of the standalone accuracy losses due to each imperfection parameter is higher than the accuracy loss when they are simultaneously present. In fact for some $P$'s, $AAL_{10}$ is greater than $SAL_{10}$ by up to 20\%, while the mean difference between the two (over all the 2000 $P$s) is $\approx$~10.5\%. Fig. \ref{allvarying}(b) highlights the median values of each parameter, $SAL_{10}$, and $AAL_{10}$ across the different $P$s in each $SAL_{10}$ band.\par


\begin{figure}[t]
 \centering
   \subfigure[]
   {\includegraphics[width=0.48\textwidth]{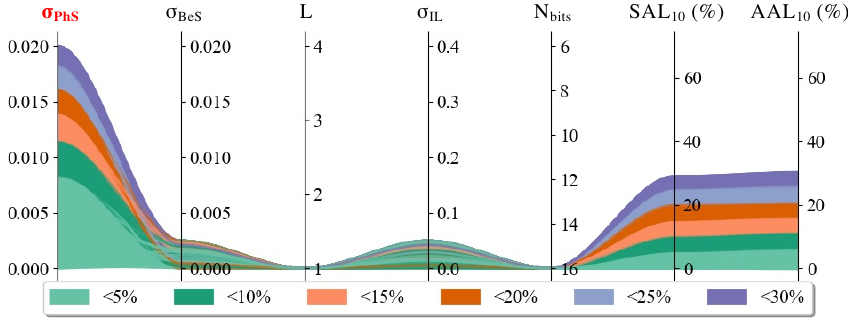}}
  \subfigure[]{
  \includegraphics[width=0.48\textwidth]{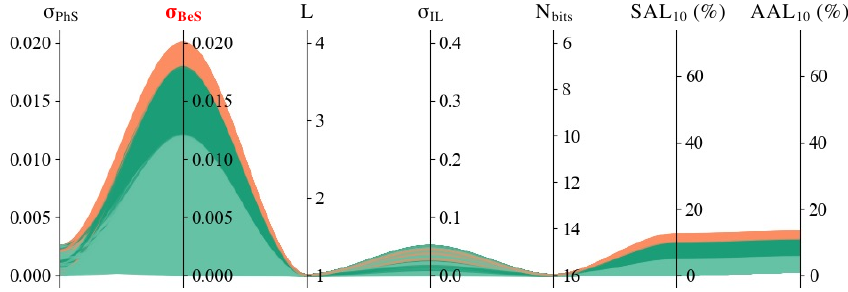} \label{ltpplots}
  }
  \subfigure[]{
  \includegraphics[width=0.48\textwidth]{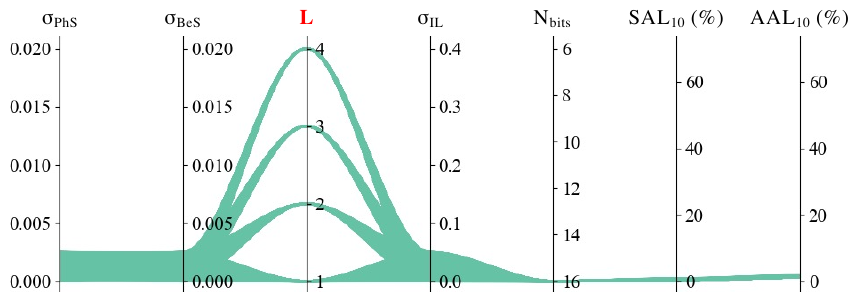} \label{ltpplots}
  }
  \subfigure[]{
  \includegraphics[width=0.48\textwidth]{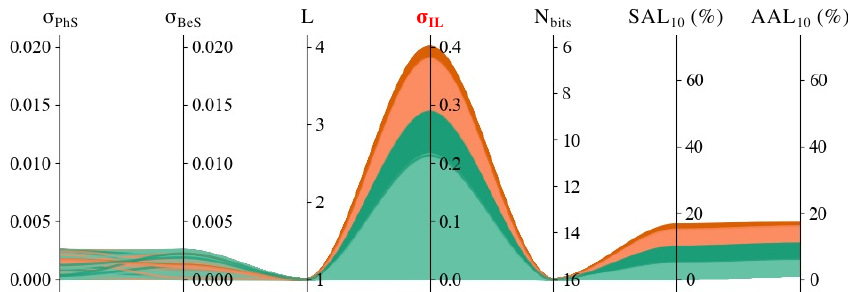} \label{ltpplots}
  }
  \subfigure[]{
  \includegraphics[width=0.48\textwidth]{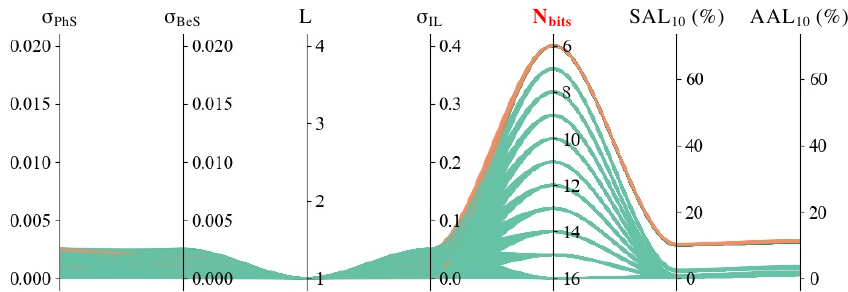} \label{ltpplots}
  }
  \vspace{-0.05in}
  \caption{$SAL_{10}$ and $AAL_{10}$ for 2000 randomly generated imperfection parameter sets where (a) $\sigma_{PhS}$, (b) $\sigma_{BeS}$, (c) $L$, (d) $\sigma_{IL}$, and (e) $N_{bits}$ individually vary over a wide range, respectively (shown in bold red font). In each case, the remaining parameters are restricted within a tolerable limit. All the subplots use the legend shown below (a).}
  \label{onevarying}
 \vspace{-0.3in}
 \end{figure}

We extend our case study further and perform additional simulations to answer the following questions (Qs) related to IPNN performance under simultaneous imperfections:
\begin{itemize}
    \item \textbf{Q1}: How does the IPNN accuracy vary under catastrophic imperfections from a single source while the imperfections from the remaining sources are restricted (but non-zero)?
    \item \textbf{Q2}: With respect to phase angles and splitting ratios, are localized imperfections with a large magnitude more critical than small correlated imperfections?
    \item \textbf{Q3}: How critical are imperfections affecting the PhS (e.g., phase angle variations and quantization errors due to low-precision phase encoding) compared to the other imperfections?
    \item \textbf{Q4}: Given a maximum tolerable accuracy loss, can we define a tolerable limit for each imperfection parameter? 
\end{itemize}
Simulation results in Fig. \ref{onevarying} \textbf{answer Q1} by showing the IPNN performance when one imperfection parameter varies over a larger range (see respective subfigures) while the other present imperfections are restricted within a tolerable limit. For better understanding, each color band spanning a 10\% accuracy range in Fig. \ref{allvarying} has been divided into two sub-bands, with the lighter shade denoting the lower 5\% in each band (see the legend in Fig. \ref{onevarying}(a)). We observe that with all the other imperfections within tolerable limits, variations in the phase angles (quantified by $\sigma_{PhS}$) alone can result in up to a 30\% loss in the inferencing accuracy. Note that this is in agreement with our earlier observation in Fig. \ref{compress_EXP1}, where the blue-dashed line shows an accuracy loss of $\approx$30\% for $\sigma_{PhS}=$~0.02. However, in $P$'s where parameters other than $\sigma_{PhS}$ are dominant, the corresponding $SAL_{10}(P)$'s are significantly lower. This shows that random variations in the phase angles have a dominant impact compared to the other imperfections. Observe also that in Fig. \ref{onevarying}, $SAL_{10}(P)\approx AAL_{10}(P)$ across all $P$'s. This is because in each case, only one of the imperfection parameters has a significant contribution to the $SAL_{10}$. \par 

Similarly, Figs. \ref{onevarying}(a)-(c), corresponding to the scenarios where $\sigma_{PhS}$, $\sigma_{BeS}$, and $L$ vary over a large range, respectively, \textbf{answer Q2}. We find that large uncorrelated ($L=$~1) variations in phase angles and splitting ratios lead to a significantly higher $SAL_{10}$ compared to tolerable but highly correlated ($L=$~4) variations. In other words, large localized uncertainties in PhS and BeS are more critical than distributed negligible uncertainties. Again, from Fig. \ref{onevarying}(e) (corresponding to $N_{bits}$), we observe that in the scenarios where other imperfection parameters are within tolerable limits, $SAL(P)\leq 10\%$ for all $P$'s with $N_{bits}\geq 7$. This is in agreement with Fig. \ref{fig10} above. \par


\begin{figure}[t]
 \centering
   \subfigure[]
   {\includegraphics[width=0.48\textwidth]{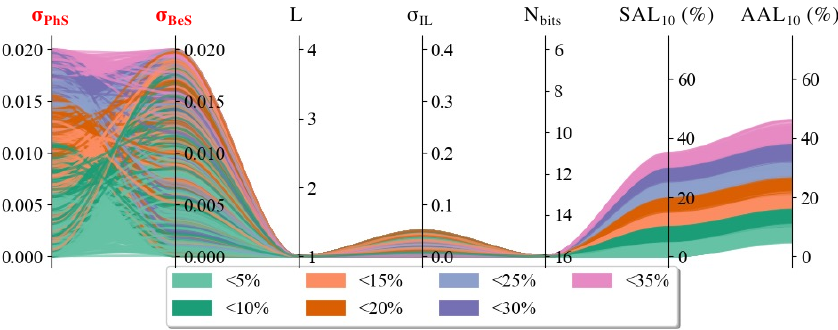}}
  \subfigure[]{
  \includegraphics[width=0.48\textwidth]{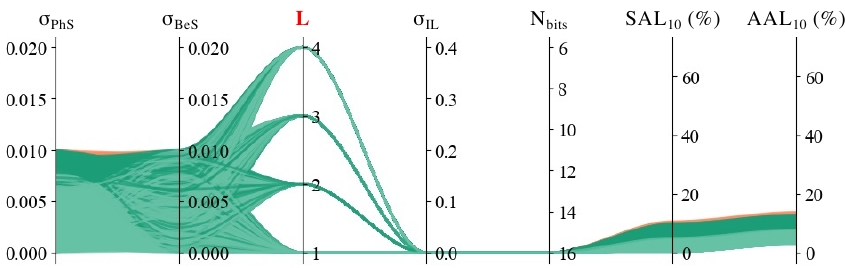}
  } \vspace{-0.05in}
  \caption{$SAL_{10}$ and $AAL_{10}$ for 2000 uncertainty parameter sets with (a) large uncorrelated and (b) small correlated variations in PhS and BeS.}
\vspace{-0.3in}
 \label{correlationvarying}
 \end{figure}


\begin{figure}[t]
 \centering
   \subfigure[]
   {\includegraphics[width=0.48\textwidth]{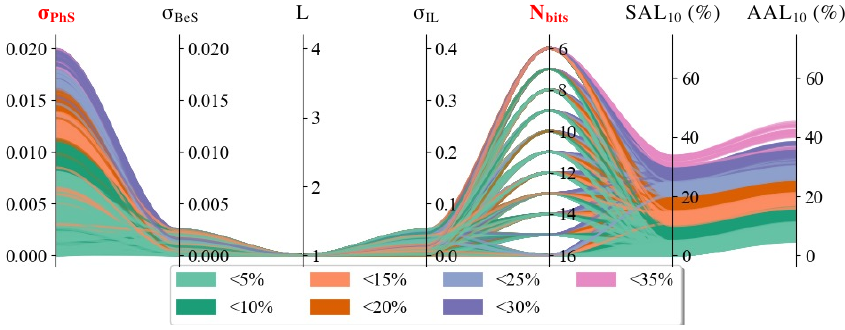}}
  \subfigure[]{
  \includegraphics[width=0.48\textwidth]{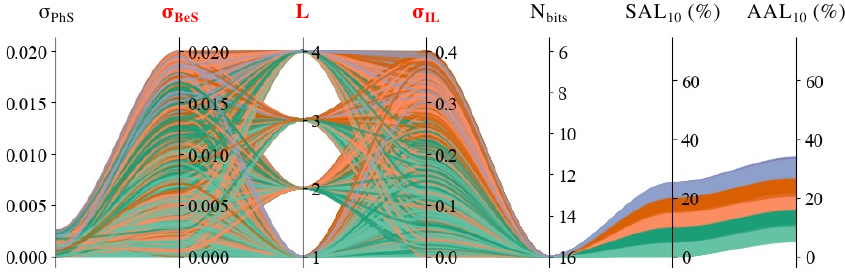}
  } 
  \subfigure[]{
  \includegraphics[width=0.48\textwidth]{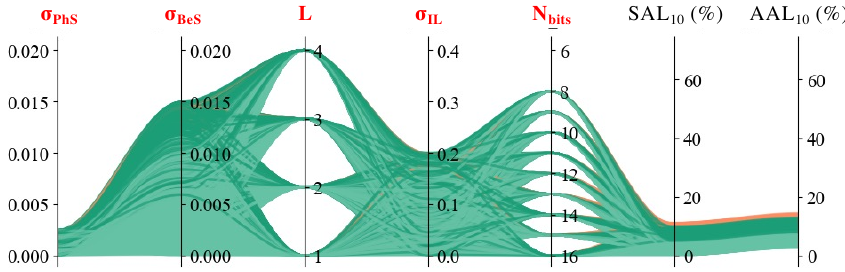}
  }
  \vspace{-0.05in}
  \caption{(a) $SAL_{10}$ and $AAL_{10}$ for 2000 randomly generated imperfection parameter sets where $\sigma_{PhS}$ and $N_{bits}$ vary over a wide range while the other parameters are restricted within a tolerable limit. (b) $SAL_{10}$ and $AAL_{10}$ for 2000 imperfection parameter sets where $\sigma_{BeS}$, $L$, and $\sigma_{IL}$) vary over a wide range while $\sigma_{PhS}$ and $N_{bits}$ are restricted within a tolerable limit. (c) Maximal imperfection parameter set $P^{*}\left(\alpha_{max}=10\%, n_p=10\right)$ obtained using trial-and-error. All the subplots use the legend shown below the top subplot.}
 \vspace{-0.2in}
 \label{phasevarying}
 \end{figure}

In Section \ref{random}, we showed that with $\sigma_{PhS}$ and $\sigma_{BeS}$ remaining \textit{constant}, uncertainties with a higher correlation length ($L$) lead to a higher inferencing accuracy loss (see Fig. \ref{fig8}). However, Fig. \ref{correlationvarying} shows that large simultaneous uncorrelated uncertainties in PhS and BeS (high $\sigma_{PhS}$ and $\sigma_{BeS}$ with low $L$) are more catastrophic than highly correlated (widespread) uncertainties of smaller magnitude (low $\sigma_{PhS}$ and $\sigma_{BeS}$ with high $L$). Uncorrelated ($L$~=1) variations with large $\sigma_{PhS}$ and $\sigma_{BeS}$ result in up to 35\% $SAL_{10}$. In contrast, most highly correlated variations ($L=$~4) with $\sigma_{PhS}=\sigma_{BeS}=$~0.01 result in a $SAL_{10}$ of only 10\%. Next, we \textbf{answer Q3}. Fig. \ref{phasevarying}(a) shows the case where the phase angles deviate significantly from their tuned values---they are prone to large localized variations ($\sigma_{PhS}$) and suffer from low-precision encoding ($N_{bits}$). We find that for such $P$'s the $SAL_{10}$ can be up to 35\%. In contrast, from Fig. \ref{phasevarying}(b) observe that if the phase variations can be restricted within a limit ($\sigma_{PhS}\leq$~0.0025) and the phase angles are encoded using $N_{bits}=$~16 bits, the maximum $SAL_{10}$ across all $P$'s is less than 25\%. To quantify the resilience of different IPNN architectures against imperfections, designers may need to find the maximum tolerable level of different imperfection parameters under which the inferencing accuracy loss remains below a threshold---a higher tolerance limit denotes a more resilient architecture. To \textbf{answer Q4}, we define the maximal imperfection parameter set for an IPNN as follows:  


\begin{definition}
    For a given IPNN, the \textbf{maximal imperfection parameter set} $P^{*}\left(\alpha_{max}, n_p\right)=\{\sigma_{PhS}^{*}, \sigma_{BeS}^{*}, L^{*}, \sigma_{IL}^{*}, N_{bits}^{*}\}$ is given by the multi-objective optimization problem:
    \begin{equation}
    \begin{aligned}
    \max_{P} \hskip0.5em\relax & \left\{\sigma_{PhS}, \sigma_{BeS}, L, \sigma_{IL}, \frac{1}{N_{bits}}\right\} & \\\textrm{s.t.} \hskip0.5em\relax & SAL_{n_p}(P)\leq \alpha_{max}.
    \end{aligned}
    \label{optim_problem}
    \end{equation}
\end{definition}
\noindent Here, $\alpha_{max}$ denotes the maximum tolerable inferencing accuracy loss, which is application dependent, and $n_p$ denotes the number of imperfection instances considered to obtain $SAL_{10}$. Note that the objective function includes the reciprocal of $N_{bits}$, as a lower $N_{bits}$ denotes a higher level of imperfections in the phase encoding. We should also mention here that $P^{*}$ is defined in a probabilistic manner. It can not be claimed, with absolute certainty, that under any imperfection instance $p$ where the parameter values are within the $P^{*}$, the IPNN accuracy loss will be less than $\alpha_{max}$. This is because, in each imperfection instance, the correlated uncertainty maps $u_{PhS}$ and $u_{BeS}$ (see Section III-D.3) and the insertion loss $IL$ are randomly sampled from a continuous Gaussian distribution. However, note that $n_p$ is crucial in this aspect. Recall that the accuracy loss over $n_p$ $p$'s are averaged to obtain $SAL_{n_p}$. Therefore, by the law of large numbers, the larger the $n_p$, the higher should be the probability that any imperfection instance with parameters within the range defined by $P^{*}$ should lead to an accuracy loss below $\alpha_{max}$. \par

Finding $P^{*}$ for a given IPNN is a non-trivial problem primarily because the constraint $SAL_{n_p}(P)\leq \alpha_{max}$ is an oracle (or black box) and does not have a closed-form expression. Additionally, the problem is mixed in nature. Indeed, $L$ and $N_{bits}$ can only take discrete values (i.e., discrete variables) while $\sigma_{PhS}$, $\sigma_{BeS}$, and $\sigma_{IL}$ are continuous in nature (i.e., continuous variables). Taking these challenges into consideration, we can use a grid-search method to approximate $P^{*}$. Fig. \ref{phasevarying}(c) shows an example where we approximate $P^{*}$ for our IPNN with $\alpha_{max}=$~10\% and $n_p=$~10. In this case, $\sigma_{PhS}^{*}=$~0.0025, $\sigma_{BeS}^{*}=$~0.015, $L=$~4, $\sigma_{IL}=$~0.2, and $N_{bits}=$~8. Evidently, these are rough estimates as we can see that there are a few $P$'s within the stipulated range, for which the $SAL_{10}$'s do exceed 10\%. In addition, the grid-search method becomes intractable as the step size decreases. Therefore, future research should focus on a computationally efficient search method for $P^{*}$.\par

%% file: appendix.tex
To realize correlated uncertainties, let us consider a $(N-1)\times 2N$ MZI grid corresponding to a $N\times N$ unitary matrix with uncertainties introduced only in the phase angles. Given $\sigma_{PhS}$, a random uncorrelated variation map for this grid is:

\begin{equation}
\begin{aligned}
    & v_{PhS}(x,y, \sigma_{PhS})\sim \mathcal{N}\left(0,4\pi^2\sigma_{PhS}^2\right) \\
    & \quad \forall~0\leq x \leq 2N-1, 0\leq y \leq N-2.
\end{aligned}
    \label{v_PhS_short}
\end{equation}
Here, $x$ and $y$ denote the coordinates of PhS, measured in units from the lower-left corner of the array (see Fig. \ref{fig6}). To model such radial variation maps, we scale the standard deviation of each cell in the grid based on its Euclidean distance from the center of the grid (coordinates (15.5, 7) for the 15$\times$32 grid). As a result, devices equidistant from the grid center will experience similar uncertainties. The random uncorrelated radial variation at any point $\left(x,y\right)$, in this case, is given by:

\begin{equation}
\begin{aligned}
v_{PhS,rad}& (x,y, \sigma_{PhS})\sim \\
& \mathcal{N}\left(0,4\pi^2\sigma_{PhS}^2 \frac{(x-\frac{2N-1}{2})^2+(y-\frac{N-2}{2})^2}{(\frac{2N-1}{2})^2+(\frac{N-2}{2})^2}\right), \\
& \forall~0\leq x \leq 2N-1, 0\leq y \leq N-2.
\end{aligned}
\label{v_PhS_long}
\end{equation}

Fig. \ref{fig7}(b) shows one such randomly generated 15$\times$32 uncorrelated radial variation map with $\sigma_{PhS}=$~0.025. The random variation map $v$ (or $v_{rad}$) is then convolved with a Gaussian kernel, $g(x,y)$, given by:
\begin{equation}
    g(x,y, L)=\frac{2}{\sqrt{\pi}L}e^{-\frac{2(x-\frac{2N-1}{2})^2+(y-\frac{N-2}{2})^2}{L^2}},
    \label{g_xyL}
\end{equation}
where $L$ denotes the correlation length of the uncertainties measured in terms of units, with one unit denoting the width of an MZI (see Fig. \ref{fig6}). Fig. \ref{fig7}(c) shows the Gaussian kernel with $L=4$ which denotes a correlation length of 4 units. The spatially correlated variation map is then given by:
\begin{equation}
    u_{PhS}(x,y, L, \sigma_{PhS})=g(x,y,L)*v(x,y,\sigma_{PhS}),
    \label{u_PhS}
\end{equation}
where $*$ denotes the convolution operation. Figs. \ref{fig7}(d)--(f) show the spatially correlated variation maps obtained when the uncorrelated variation map (Fig. \ref{fig7}(a)) is convoluted with Gaussian kernels with $L=$~2, 4, and 8 units, respectively. Observe that as $L$ increases, the uncertainties in the variation maps are spread out over larger contiguous areas. Similarly, Fig. \ref{fig7}(g) shows the correlated variation map when the uncorrelated radial variation map (Fig. \ref{fig7} (b)) is convoluted with a Gaussian kernel with $L=4$. Note that the correlated variation map for the splitting ratios, $u_{BeS}(x,y, L, \sigma_{BeS})$ is similarly obtained using the random uncorrelated variation maps given by: 
\begin{equation}
    v_{BeS}(x,y, \sigma_{BeS})\sim \mathcal{N}\left(0,\sigma_{BeS}^2/2\right),
    \label{v_BeS_short}
\end{equation}


\begin{equation}
\begin{aligned}
v_{BeS,rad}& (x,y, \sigma_{BeS}) \\
& \hspace{-1em} \sim \mathcal{N}\left(0,\frac{\sigma_{BeS}^2}{2} \frac{(x-\frac{2N-1}{2})^2+(y-\frac{N-2}{2})^2}{(\frac{2N-1}{2})^2+(\frac{N-2}{2})^2}\right).
\end{aligned}
\label{v_BeS_long}
\end{equation} 